\documentclass[twocolumn,linenumbers]{emulateapj}
\usepackage{amsmath}
\usepackage{nicefrac}
\usepackage[none]{hyphenat}
\usepackage[T1]{fontenc}
\usepackage{graphicx}
\usepackage{hyperref}
\usepackage{verbatimbox}
\usepackage{lineno}
\usepackage{booktabs}

\def\Msun{M$_{\odot}$}

\def\deg{$^{\circ}$}
\usepackage{adjustbox}

\begin{document}

\title{Spin-Orbit Alignment of Early-type Astrometric Binaries
\\and the Origin of Slow Rotators}

\author{Chase L. Smith\altaffilmark{1}, Maxwell Moe\altaffilmark{1}, and Kaitlin~M.~Kratter\altaffilmark{2}}

\altaffiltext{1}{Department of Physics and Astronomy, University of Wyoming, 1000 E.~University~Ave., Dept.~3905, Laramie, WY, 82071, USA; csmit151@uwyo.edu, mmoe2@uwyo.edu}

\altaffiltext{2}{Steward Observatory, University of Arizona, 933 N. Cherry Ave., Tucson, AZ 85721, USA}


\begin{abstract}
The spin-orbit alignment of binary stars traces their formation and accretion history. Previous studies of spin-orbit alignment have been limited to small samples, slowly rotating solar-type stars, and/or wide visual binaries that not surprisingly manifest random spin-orbit orientations. We analyze 917 {\it Gaia} astrometric binaries across periods $P$ = 100\,-\,3,000 days ($a$ = 0.5\,-\,5~au) that have B8-F1\,IV/V primaries ($M_1$ = 1.5\,-\,3~\Msun) and measured projected rotational velocities $v$\,sin\,$i$. The primary stars in face-on orbits exhibit substantially smaller $v$\,sin\,$i$ compared to those in edge-on orbits at the 6$\sigma$ level, demonstrating significant spin-orbit alignment. The primaries in our astrometric binaries are rotating more slowly than their single-star or wide-binary counterparts and therefore comprise the slow-rotator population in the observed bimodal rotational velocity distribution of early-type stars. We discuss formation models of close binaries where some of the disk angular momentum is transferred to the orbit and/or secondary spin, quenching angular momentum flow to the primary spin. The primaries in astrometric binaries with small mass ratios $q$~=~$M_2$/$M_1$~$<$~0.3 possess even smaller $v$\,sin\,$i$, consistent with model predictions. Meanwhile, astrometric binaries with large eccentricities $e$~$>$~0.4 do not display spin-orbit alignment or spin reduction. Using a Monte Carlo technique, we measure a spin-orbit alignment fraction of $F_{\rm align}$ = 75\%\,$\pm$\,5\% and an average spin reduction factor of $\langle S_{\rm align} \rangle$ = 0.43\,$\pm$\,0.04. We conclude that 75\% of close A-type binaries likely experienced circumbinary disk accretion and probably formed via disk fragmentation and inward disk migration. The remaining 25\%, mostly those with $e$~$>$~0.4, likely formed via core fragmentation and orbital decay via dynamical friction. \vspace*{0.4cm}
\end{abstract}


\section{Introduction}
\label{sec:Intro}

Most stars are born in binary or multiple systems \citep{Offner2023}. Close binary stars ($a$ $<$ 10~au) can interact via mass transfer and are the progenitors of many different types of exotic stars and transients \citep{DeMarco&Izzard2017}. It is therefore important to characterize the physical properties of close binaries in order to better constrain their formation processes and subsequent evolution. 

Close binaries form through two main channels: disk fragmentation and core/filament fragmentation (see \citealt{Offner2023} for a recent review). For the former, a massive protostellar disk can become gravitationally unstable and fragment, initially forming a brown dwarf companion in the cool outer disk across $a$~=~20\,-\,300~au \citep{Kratter2008,Kratter&Lodato2016}. The direction and timescale of orbital migration of binaries embedded in disks depend on many uncertain variables (see \citealt{Lai2023} for a recent review). Nonetheless, a brown dwarf companion that recently fragmented from a massive disk is expected to migrate inward to within $a$~$<$~10~au on rapid viscous timescales. The close binary subsequently clears out an inner cavity and accretes from a circumbinary disk 
\citep{Artymowicz1994,Artymowicz1996,Clarke1996}. Most of the infalling material from the circumbinary disk is directed toward and accreted by the secondary \citep{Farris2014,Young2015}, driving the mass ratio $q$ = $M_2$/$M_1$ toward unity. \citet{Tokovinin2020} developed a toy model of disk fragmentation, migration, and accretion that reproduces many observed features of solar-type binaries within $a$~$<$~1 au and OB binaries with $a$ $<$ 10 au. At wider separations, however, their model underestimates the observed binary frequency, suggesting a different formation mechanism begins to dominate.

On larger scales beyond $a$~$>$~300~au, molecular cores and filaments can fragment into binaries that may subsequently decay to shorter separations via dynamical friction or disk capture \citep{Bate2002,Bate2012,Offner2012,Munoz2015,Lee2019}. Once the binary migrates to within $a$~$\lesssim$~30~au, the system can undergo the same processes of disk migration and circumbinary disk accretion as binaries that initially formed via disk fragmentation. However, migration of core fragmentation binaries toward close separations $a$~$<$~10~au likely occurs when the disk is less massive, on average, compared to those that initially formed from a gravitationally unstable disk. In particular, the protostellar disk lifetimes of Herbig Ae/Be stars ($M$~$>$~2\,\Msun) are only 1~Myr \citep{Hernandez2005,Boissier2011,Ribas2015,Grant2023}, and thus only the subset of core-fragmentation binaries that can migrate to short separations on rapid timescales will likely accrete substantial material from a circumbinary disk. Close early-type binaries that derive from the two different formation channels likely experience different levels of circumbinary disk accretion and thus probably exhibit different observational signatures.


The spin-orbit alignment of a binary traces how the system gained and transferred angular momentum during its formation. Hydrodynamic simulations demonstrate that the majority of close binaries have some degree of spin-orbit alignment, regardless of their formation mechanism \citep{Bate2010,Bate2012,Bate2018}. Nonetheless, binaries that formed on initially smaller disk scales have both smaller spin-orbit angles and smaller eccentricities. For example, of the 21 close binaries with final separations $a_{\rm f}$~$<$~10~au in the \citet{Bate2012} simulation (see their Table~3), 11 initially fragmented on large core scales beyond $a_{\rm i}$~$>$~300~au. They ended with an average eccentricity of $\langle e \rangle$ = 0.55\,$\pm$\,0.06 and average inclination between primary spin and orbit of $\langle i \rangle$ = 62\deg\,$\pm$\,10\deg. The majority (8/11 = 73\%) of these core-fragmentation systems that end as close binaries prefer prograde orbits relative to their spins. Meanwhile, the remaining 10 close binaries that initially fragmented on small disk scales of $a_{\rm i}$ = 20\,-\,230 au ended with substantially smaller eccentricities $\langle e \rangle$ = 0.32\,$\pm$\,0.06 and spin-orbit inclinations $\langle i \rangle$ = 31\deg\,$\pm$\,7\deg\ (100\% prograde). 

The majority of the close binaries simulated in \citet{Bate2012} are low mass that accreted from long-lived circumbinary disks that could easily realign the spins to the binary orbits. If we instead limit the sample to the 10 binaries with $M_1$ $>$ 1\,\Msun\ and $a$ $<$ 10~au from \citet{Bate2012} and supplement with the 11 binaries within the same parameter space from the solar-metallicity simulation of \citet{Bate2019}, then the predicted degree of spin-orbit alignment is substantially weaker. For the subset with $e$ $>$ 0.5, the average spin-orbit inclination is $\langle i \rangle$ = 77\deg\,$\pm$\,13\deg, only marginally different from isotropic orientations. For those with $e$~$<$~0.5, the average spin-orbit inclination remains rather large at $\langle i \rangle$ = 48\deg\,$\pm$\,10\deg. For close early-type binaries, accretion from the short-lived circumbinary disks can only moderately dampen the eccentricities and slightly reduce the spin-orbit angles. In this regime, mass ratio seems to better correlate with spin-orbit alignment than eccentricity. The subset with $q$ $>$ 0.6 has $\langle i \rangle$ = 71\deg\,$\pm$\,10\deg\ while those with $q$ $<$ 0.6 have $\langle i \rangle$ = 28\deg\,$\pm$\,10\deg. Disk fragmentation naturally produces binaries with well aligned spins and low-mass companions.  Obviously, the degree of spin-orbit alignment and circumbinary disk accretion depend on a complex interplay of many uncertain factors, motivating the need to quantitatively measure the fraction of close binaries that exhibit spin-orbit alignment.


We organize the rest of this paper as follows. In section~\ref{sec:Review}, we first summarize previous measurements of spin-orbit alignment and share how our sample of early-type astrometric binaries is specifically designed to overcome previous shortcomings. We compute in section~\ref{sec:APRV} the average projected rotational velocities for all stars as a function of spectral type and luminosity class. In section~\ref{sec:AstroBin}, we discuss our selection of astrometric binaries from {\it Gaia} DR3, their degree of spin-orbit alignment, and other observed properties of our sample. We highlight our serendipitous discovery of slow rotators in section~\ref{sec:SlowRotators}. In section~\ref{sec:Dep}, we compare how spin-orbit alignment and the slow-rotator fraction varies as a function of binary orbital parameters. In section \ref{sec:MC}, we conduct a Monte Carlo population synthesis study to measure the spin-orbit alignment and rotational velocity distribution of our sample. We conclude in section~\ref{sec:Conclusions}.

\section{Review of Spin-Orbit Alignment Studies}
\label{sec:Review}

\subsection{Visual Binaries}

Two early studies showed that early-type visual binaries beyond $a$~$>$~100~au \citep{Slettebak1963} and $a$~$>$~40~au \citep{Levato1974} do not show any significant spin-orbit alignment, as expected for such wide systems. \citet{Weis1974} subsequently compiled projected rotational velocities $v$\,sin\,$i$ and orbital inclinations $i_{\rm orb}$ for 132 early-type visual binaries, mostly across $a$ = 10\,-\,100~au. For each binary, they computed the fractionalized projected rotational velocity $\langle V_f \rangle$ as the ratio of the projected rotational velocity of the primary star versus the mean projected rotational velocity of all other stars of matching spectral type and luminosity class. Their subset of 46 F-type visual binaries exhibited some degree of spin-orbit alignment whereby the primaries in more face-on orbits tended to have smaller $\langle V_f \rangle$. However, their F-type sample included only five visual binaries with such face-on orbits below sin\,$i_{\rm orb}$~$<$~0.6.  Moreover, \citet{Weis1974} did not find any significant spin-orbit alignment for their subset of A-type visual binaries. They examined the degree of spin-orbit alignment as a function of orbital period, but could not detect any significant trend. Clearly, a larger sample of close early-type binaries is needed to detect the expectedly larger degree of spin-orbit alignment at smaller separations.

Two decades later, \cite{Hale1994} measured the spin-orbit alignment of 45 solar-type visual binaries. Unlike the previous studies that compared $v$\,sin\,$i$ versus $i_{\rm orb}$ to measure the degree of spin-orbit alignment in a statistical sense, \citet{Hale1994} measured stellar radii, rotation periods, and $v$\,sin\,$i$ to compute the actual inclination difference $\Delta i$ = |$i_{\rm spin}$\,$-$\,$i_{\rm orb}$| between stellar spin and orbital inclination for each system. They concluded that wide solar-type visual binaries beyond $a$~$>$~30~au exhibit sufficiently large $\Delta i$ to be consistent with random spin-orbit orientations. Conversely, their 15 solar-type binaries within $a$ $<$ 20~au all had small inclination differences below $\Delta i$ $<$ 20\deg, suggesting significant spin-orbit alignment. 

However, in a recent follow-up analysis, \citet{Justesen2020} more accurately measured the stellar radii from {\it Gaia} parallaxes and $v$\,sin\,$i$ from their own high-resolution spectra for a subset of the \citet{Hale1994} sample of solar-type visual binaries. Their updated measurements revealed that some of the binaries below $a$ $<$ 20~au actually have larger inclination differences, several as high as $\Delta i$ = 40\deg. \citet{Justesen2020} concluded that their subset of close solar-type visual binaries was insufficient to discern any statistically significant spin-orbit alignment.

\subsection{Spectroscopic Binaries}

The Binaries Are Not Always Neatly Aligned (BANANA) project has measured obliquities of very close spectroscopic binaries \citep{Marcussen2022}. In their sample of 43 early-type and solar-type spectroscopic binaries, 40 are consistent with alignment. However, a significant majority of their spectroscopic binaries have very short orbital periods $P$ $<$ 10 days, and so it is difficult to discern whether alignment traces their formation pathway or is due to tidal interactions during the main-sequence (MS) or pre-MS. 

\subsection{Motivating our Sample of Astrometric Binaries}

With the release of {\it Gaia} DR3, orbital parameters of unprecedented accuracy are available for $>$10$^{5}$ astrometric binaries spanning a wide range of spectral types \citep{Halbwachs2023}. In this study, we examine the spin-orbit alignment of early-type astrometric binaries in {\it Gaia} DR3, which allows us to overcome three shortcomings of the previous studies. First, we limit our sample to astrometric binaries across $P$~=~100\,-\,3,000~days ($a$ = 0.5\,-\,5 au), wide enough where tides have negligibly affected their spins but close enough where we expect significant spin-orbit alignment. Our astrometric binary sample effectively fills in the gap between spectroscopic and visual binaries. Second, early-type primaries above the \citet{Kraft1967} break have substantially larger projected rotation velocities $v$\,sin\,$i$ = 30\,-\,300 km~s$^{-1}$, which provides superior leverage in constraining their spin-orbit alignment. Unlike solar-type primaries with $v$\,sin\,$i$ = 2 - 10 km~s$^{-1}$, where small measurement errors can lead to large biases in spin-orbit alignment, even low-resolution spectra of early-type stars yield reliable $v$\,sin\,$i$ measurements. Finally, our final sample is more than an order of magnitude larger than previous surveys of spin-orbit alignment. Similar to \cite{Weis1974}, we can simply analyze the distribution of $v$\,sin\,$i$ as a function of orbital inclination $i_{\rm orb}$ rather than performing the complex task of measuring stellar radii and rotation periods.



\section{Average Projected Rotational Velocities}
\label{sec:APRV}

Stellar rotational velocities increase dramatically above the Kraft break \citep[$T_{\rm eff}$~$>$~6,250~K, $<$\,F7\,V, $M$~$>$~1.25\,\Msun;][]{Kraft1967, Glebocki2005}. Assuming random orientations and using an iterative technique, \citet{Royer2007} and \citet{Zorec2012} reconstructed the true rotational velocity $v_{\rm rot}$ distributions based on the observed projected rotational velocity $v$\,sin\,$i$ distributions. For early-F/late-A dwarfs (1.5\,-\,2.0\,\Msun), they recovered a mostly unimodal Maxwellian distribution that peaks near $v_{\rm rot}$~=~140~km~s$^{-1}$. Meanwhile, for late-B/early-A dwarfs (2\,-\,4\,\Msun), \citet{Royer2007} and \citet{Zorec2012} measured a bimodal velocity distribution where 80\%\,-\,95\% follow a broad Maxwellian distribution that peaks near $v_{\rm rot}$~=~190~km~s$^{-1}$. The remaining 5\%\,-\,20\% comprise the "slow-rotator" population that peaks near $v_{\rm rot}$~=~50~km~s$^{-1}$. \citet{Dufton2013} subsequently showed that early-B stars (8\,-\,15\,\Msun) also follow a bimodal distribution with a slightly larger slow-rotator population of 25\% that spans $v_{\rm rot}$ = 0\,-\,100~km~s$^{-1}$. The remaining 75\% of early-B stars have fast rotational velocities that broadly peak near $v_{\rm rot}$ = 260 km~s$^{-1}$. With increasing stellar mass across $M$ = 1.5\,-\,15\,\Msun, the average rotational velocity increases despite the increase in the slow-rotator fraction. 

Similar to \citet{Weis1974}, we determine the mean projected rotational velocity $\langle v$\,sin\,$i\rangle_{\rm ST}$ as a function of spectral type and luminosity class. We eventually compute the ratio:

\begin{equation}
r_{\rm vsini} = \frac{v\,{\rm sin}\,i}{\langle v\,{\rm sin}\,i\rangle_{\rm ST}}
\label{eqn:rvsini}
\end{equation}

\noindent between the projected rotational velocity of an individual primary star in a binary and the corresponding average $\langle v\,{\rm sin}\,i\rangle_{\rm ST}$ of all stars given the same spectral type and luminosity class (see section \ref{sec:AstroBin}). The parameter $r_{\rm vsini}$ is a more accurate indicator of spin orientation than $v$\,sin\,$i$ alone because it accounts for the variations in rotational velocity with respect to stellar mass and age. Small values of $r_{\rm vsini}$~$\lesssim$~0.2 suggest pole-on spins while large values $\gtrsim$\,2 suggest edge-on orientations. A population of primary stars with random spin orientations and the same rotational velocity distribution as all stars will have an average ratio of $\langle r_{\rm vsini} \rangle$ = 1 by definition.

\begin{figure}[t!]
\includegraphics[scale = 0.61]{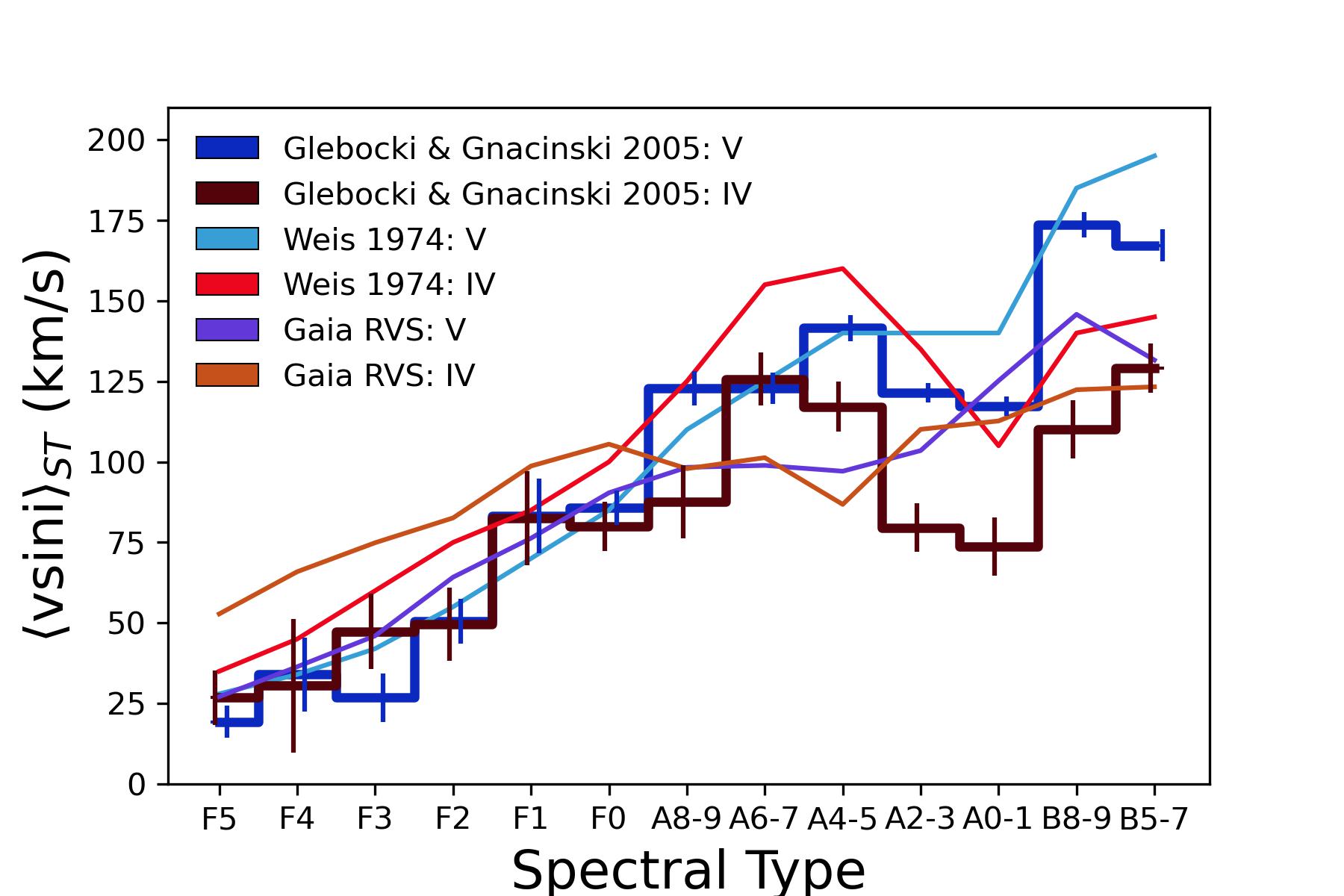}
\caption{Average projected rotational velocities $\langle v$\,sin\,$i\rangle_{\rm ST}$ of all stars (single and binary) as a function of spectral type and luminosity class. Bluish and reddish colors correspond to dwarfs and subgiants, respectively. We compare three different surveys: \citet[][thick dark blue and dark red]{Glebocki2005},  \citet[][dashed light blue and light red]{Weis1974}, and {\it Gaia} RVS (thin purple and brown).} 
\vspace*{4mm}
\label{fig:meanvsini}
\end{figure}

In Fig.~\ref{fig:meanvsini}, we plot $\langle v$\,sin\,$i\rangle_{\rm ST}$ from the full \citet{Weis1974} sample for spectral types B5\,-\,F5 and luminosity classes IV and V. \citet{Glebocki2005} subsequently synthesized rotational velocities from various surveys and computed weighted averages for individual systems. Some of our adopted $v$\,sin\,$i$ measurements derive from their weighted means. We display $\langle v$\,sin\,$i\rangle_{\rm ST}$ and the standard deviations of those means from the \citet{Glebocki2005} catalog in Fig.~\ref{fig:meanvsini}. Many of the early-type primaries in our astrometric binaries have velocity broadening functions measured by {\it Gaia}'s Radial Velocity Spectrometer \citep[RVS;][]{GaiaDR3}. \citet{Fremat2023} recently demonstrated that the {\it Gaia} RVS broadening parameter $v_{\rm broad}$ closely matches $v$\,sin\,$i$ for early-type stars brighter than G~$\lesssim$~10 (see their Fig.~4). The main deviation occurs for pole-on orientations where the velocity width is dominated by microturbulence and thermal broadening $v_{\rm t}$ $\approx$ 6~km~s$^{-1}$. Using the {\it Gaia} archive\footnote{https://gea.esac.esa.int/archive/}, we select the 27,468 stars brighter than G~$<$~9.5 that have parallaxes $\varpi$~$>$~2~mas (distances $d$ = 1/$\varpi$ $<$ 500~pc), measured $v_{\rm broad}$, and General Stellar Parameters from Photometry (GSP-Phot) effective temperatures $T_{\rm eff}$~=~6,450\,-\,16,000\,K and surface gravities log\,$g$~=~3.3\,-\,4.6. We adopt the empirical relations in \citet{Pecaut2013} to map $T_{\rm eff}$ to spectral type, and we designate early-type stars with log\,$g$ = 3.8\,-\,4.6 as dwarfs and those with log\,$g$ = 3.3\,-\,3.8 as subgiants. To compute the projected rotational velocity, we subtract the average thermal velocity $\langle v_{\rm t} \rangle$ = 6~km~s$^{-1}$ from the {\it Gaia} broadening parameter in quadrature:

\begin{equation}
v\,{\rm sin}\,i = \big({\rm max}\{v_{\rm broad},\langle v_{\rm t} \rangle\}^2 - \langle v_{\rm t} \rangle^2 \big)^{\nicefrac{1}{2}},
\label{eqn:Gaiavsini}
\end{equation}

\noindent where we set $v$\,sin\,$i$ = 0~km~s$^{-1}$ in the rare instances that $v_{\rm broad}$ $<$ 6~km~s$^{-1}$. In Fig.~\ref{fig:meanvsini}, we display $\langle v$\,sin\,$i\rangle_{\rm ST}$ based on {\it Gaia} RVS for the same bins of spectral type and luminosity class. 

We see several consistent trends among the three sets of $\langle v$\,sin\,$i\rangle_{\rm ST}$ distributions in Fig.~\ref{fig:meanvsini}. For example, the mean projected rotational velocity quickly increases above the Kraft break from $\langle v$\,sin\,$i\rangle_{\rm ST}$ = 30\,km\,s$^{-1}$ for F5 dwarfs (6,550\,K) to 80\,km\,s$^{-1}$ for F0 dwarfs (7,220\,K). F-type subgiants rotate more rapidly than their dwarf counterparts because subgiants evolved from slightly more massive progenitors. Meanwhile, early-A/late-B dwarfs rotate more quickly than their corresponding subgiants. There are also some quantitative differences. Most notably, the {\it Gaia} RVS broadening parameter for mid-F subgiants is substantially larger { than the ground-based \citet{Weis1974} and \citet{Glebocki2005} measurements}, most likely due to the systematic errors when mapping the GSP-Phot $T_{\rm eff}$ and log\,$g$ parameters to spectral type and luminosity class. In any case, mid-F stars rotate more slowly than late-B/A stars and thus offer less leverage in determining their spin axes. We subsequently consider only primary stars with B8-F1\,IV-V primaries ($T_{\rm eff}$ = 6,885\,-\,12,500\,K).

In Table~\ref{table:Averagevsini}, we list $\langle v$\,sin\,$i\rangle_{\rm ST}$ from \citet{Weis1974}, \citet{Glebocki2005}, and {\it Gaia} RVS for our adopted range of spectral types. We also utilize VizieR\footnote{https://vizier.unistra.fr/} to query $v$\,sin\,$i$ from two additional spectroscopic surveys: the 6$^{\rm th}$ data release of LAMOST \citep{Xiang2022} and the 6$^{\rm th}$ data release of RAVE \citep{Steinmetz2020}. For LAMOST (Vizier catalog J/A+A/662/A66/table5), we select stars brighter than $G$~$<$~11 that span the effective temperatures and surface gravities for each spectral type according to the \citet{Pecaut2013} relations. In Table~\ref{table:Averagevsini}, we report the LAMOST $\langle v$\,sin\,$i\rangle_{\rm ST}$ values, which are quite similar to the {\it Gaia} RVS averages.  For RAVE, we select effective temperatures and surface gravities from their Bayesian Distances, Ages, and Stellar Parameters catalog (III/283/bdasp on VizieR). We cross-match each spectral type sample with their SPARV pipeline parameter catalog (III/283/aux) for $v$\,sin\,$i$ based on spectra with signal-to-noise ratios SNR~$>$~12. The RAVE stellar parameters become unreliable for B/early-A stars, and so we report the RAVE $\langle v$\,sin\,$i\rangle_{\rm ST}$ values for only mid-A to F1 stars in Table~\ref{table:Averagevsini}. The RAVE results are quite similar to the \citet{Weis1974} average projected velocities. In the final column of Table~\ref{table:Averagevsini}, we compute the average $\langle v$\,sin\,$i\rangle_{\rm ST}$ across the five different surveys for each spectral type. The deviations of any individual spectroscopic survey from the overall average are typically only 5\%\,-\,20\%.


\renewcommand{\arraystretch}{1.3}
\setlength{\tabcolsep}{3.5pt}
\begin{deluxetable}{ccccccc}[b!]
\tabletypesize{\footnotesize}
\tablecaption{Mean projected rotational velocities $\langle v$\,sin\,$i\rangle_{\rm ST}$ in km~s$^{-1}$}
\startdata
  & Weis74 & G\&G05 & {\it Gaia} & LAMOST & RAVE & Average \\
\hline
F1\,V &70 &83.2 &76.3 &73.6 &73.7 &75 \\
F0\,V &85 &85.8 &90.4 &82.2 &91.1 &87 \\
A8/9\,V &110 &122.9 &98.3 &90.1 &97.8 &104 \\
A6/7\,V &125 &122.8 &98.9 &92.1 &115.1 &111 \\
A4/5\,V &140 &141.5 &97.1 &100.4 &118.6 &120 \\
A2/3\,V &140 &121.4 &103.5 &108.2 &137.1 &122 \\
A0/1\,V &140 &117.3 &125.2 &124.8 &- &127 \\
B8/9\,V &185 &173.6 &145.8 &142.6 &- &162 \\
\hline
F1\,IV &85 &82.5 &98.7 &107.3 &85.0 &92 \\
F0\,IV &100 &80 &105.5 &121.4 &105.3 &102 \\
A8/9\,IV &125 &87.6 &97.9 &128.3 &111.9 &110 \\
A6/7\,IV &155 &125.7 &101.3 &129.6 &136.4 &130 \\
A4/5\,IV &160 &117.2 &86.8 &129.1 &159.4 &131 \\
A2/3\,IV &135 &79.5 &101.1 &120.7 &- &109 \\
A0/1\,IV &105 &73.7 &112.7 &122.0 &- &103 \\
B8/9\,IV &140 &110.2 &122.4 &131.9 &- &126
\enddata
 \tablecomments{Dwarfs (top) and subgiants (bottom) from \citet{Weis1974}, \citet{Glebocki2005}, {\it Gaia} RVS, LAMOST \citep{Xiang2022} and RAVE \citet{Steinmetz2020}. \vspace*{2mm}}
 \label{table:Averagevsini}
\end{deluxetable}
\setlength{\tabcolsep}{6pt}
\renewcommand{\arraystretch}{1.0}


\section{Astrometric Binaries}
\label{sec:AstroBin}
\subsection{Sample Selection}

{\it Gaia} DR3 released orbital parameters for 165,500 astrometric binaries spanning a wide range of spectral types and concentrated across periods $P$~=~80\,-\,2,000~days \citep{Halbwachs2023}. Using the {\it Gaia} archive, we initially select the 2,093 astrometric binaries within $d$~$<$~500~pc ($\varpi$~$>$~2~mas) that have orbital periods beyond $P$~$>$~100~days, which ensures that tides have negligibly affected their stellar spins, and photometric parameters $-$2.0~$<$~M$_{\rm G}$~$<$~3.1 and $-$0.2~$<$~$G_{\rm BP}$\,$-$\,$G_{\rm RP}$~$<$~0.6, which encompass B8\,-\,F1\,IV/V stars with zero to moderate dust reddening \citep{Pecaut2013}. For each astrometric binary, the {\it Gaia} archive lists its orbital period $P$, eccentricity $e$, and orbital Thieles-Innes elements A, B, F and G, which describe the elliptical motion of the primary star relative to the photocenter  \citep{Binnendijk1960,Heintz1978,Halbwachs2023}. We compute the orbital inclination $i_{\rm orb}$ and { astrometric binary} mass function $f_{\rm M}$ from the Thieles-Innes elements and other parameters as described in \citet{Halbwachs2023}, specifically their Eqn.~13 and Appendix~A Eqns.~A.2\,-\,A.6. 

As further described below, we cross-match our astrometric binary sample with Simbad\footnote{http://simbad.u-strasbg.fr/simbad/} and six different ground-based spectroscopic surveys. We remove the 965 systems with spectral types and/or luminosity classes that are unknown or outside our adopted interval. Of the 2,093 $-$ 965 = 1,128 astrometric binaries with B8-F1\,IV/V primaries, we keep the 917 (81\%) that have {\it Gaia} $v_{\rm broad}$ and/or ground-based $v$\,sin\,$i$ as our final sample. In Table~A1, we list the relevant {\it Gaia} parameters, Simbad name, and ground-based spectroscopic information for our final sample (first hour of RA in printed version; table of all 917 objects available electronically). 

For the 852 systems with {\it Gaia} $v_{\rm broad}$ measurements, we compute $v$\,sin\,$i$ according to Eqn.~\ref{eqn:Gaiavsini}. Of these, 785 have {\it Gaia} GSP-Phot $T_{\rm eff}$ and log\,$g$ parameters, and in Table~A1 we list their corresponding spectral types and luminosity classes. For 36 systems with $v_{\rm broad}$ but without GSP-Phot parameters, we adopt the ground-based measurements of spectral type and luminosity class. We adopt the Simbad spectral type and luminosity class for the 31 remaining systems with $v_{\rm broad}$ that have neither {\it Gaia} nor ground-based measurements of $T_{\rm eff}$ and log\,$g$. For all 852 systems with {\it Gaia} $v_{\rm broad}$, we adopt the corresponding $\langle v$\,sin\,$i\rangle_{\rm ST}$
from the {\it Gaia} column in Table~\ref{table:Averagevsini} and compute $r_{\rm vsini}$ from Eqn.~\ref{eqn:rvsini} accordingly.

\renewcommand{\arraystretch}{1.5}
\setlength{\tabcolsep}{4.5pt}
\begin{deluxetable}{lccccc}
\tabletypesize{\footnotesize}
\tablecaption{Differences between {\it Gaia} and Ground-based $v$\,sin\,$i$}
\startdata
 & & \multicolumn{2}{c}{$\Delta v$\,sin\,$i$ (km~s$^{-1}$)} & \multicolumn{2}{c}{$\Delta $r$_{\rm vsini}$}
 \vspace*{-0.1cm} \\
 Survey & N & $\mu$ & $\sigma$& $\mu$ &  $\sigma$\\
\hline
G\&G05 & 4 & $-$11\,$\pm$\,12 & 24 & $-$0.07\,$\pm$\,0.10 & 0.19 \\
LAMOST & 52 & 6\,$\pm$\,5 & 36 & 0.05\,$\pm$\,0.05 & 0.36 \\
RAVE & 105 & $-$17\,$\pm$\,1 & 11 & $-$0.24\,$\pm$\,0.02 & 0.16 \\
LAMOST-MRS & 8 & $-$1\,$\pm$\,4 & 11 & 0.18\,$\pm$\,0.06 & 0.17 \\
GALAH &  20 & 1\,$\pm$\,3 & 11 & 0.01\,$\pm$\,0.03 & 0.13 \\
APOGEE & 18 & $-$8\,$\pm$\,2 & 8 & $-$0.07\,$\pm$\,0.02 & 0.09 \\
\hline
Total & 207 & $-$8\,$\pm$\,2 & 23 & $-$0.11\,$\pm$\,0.02 & 0.26 \\
G $<$ 9.5 & 37 & $-$6\,$\pm\,$3 & 18 & $-$0.08\,$\pm$\,0.04 & 0.23
\enddata
 \tablecomments{We report the number N of astrometric binaries common to both samples and the corresponding mean offsets and rms residuals between {\it Gaia} versus the six ground-based surveys: \citet{Glebocki2005}, LAMOST \citet{Xiang2022}, RAVE \citet{Steinmetz2020}, LAMOST-RMS \citet{Sun2021}, GALAH \citet{Buder2021}, and APOGEE \citet{Jonsson2020}.    }
 \label{Tab:Gaia_vs_ground}
\end{deluxetable}
\setlength{\tabcolsep}{6pt}
\renewcommand{\arraystretch}{1.0}

We next describe our procedure for synthesizing $v$\,sin\,$i$ and $r_{\rm vsini}$ from the six ground-based spectroscopic catalogs. We summarize the subset that also have {\it Gaia} $v_{\rm broad}$ values in Table~\ref{Tab:Gaia_vs_ground}. \citet{Glebocki2005} lists the spectral types, luminosity classes, and $v$\,sin\,$i$ for 12 astrometric binaries in our final sample, of which 4 also have {\it Gaia} $v_{\rm broad}$ measurements. We adopt the corresponding $\langle v$\,sin\,$i\rangle_{\rm ST}$
from the G\&G05 column in Table~\ref{table:Averagevsini} to compute $r_{\rm vsini}$. Similarly, 63 astrometric binaries are included in the LAMOST hot star catalog \citep{Xiang2022}, of which 52 also have {\it Gaia} $v_{\rm broad}$ values, and we incorporate the corresponding $\langle v$\,sin\,$i\rangle_{\rm ST}$
from the LAMOST column in Table~\ref{table:Averagevsini}. We cross-match our astrometric binary sample with the RAVE \citep{Steinmetz2020} Bayesian Distances, Ages, and Stellar Parameters catalog (III/283/bdasp on VizieR) if available and otherwise their MADERA stellar parameters (III/283/madera). We include 135 astrometric binaries with measured $v$\,sin\,$i$ and spectral SNR $>$ 12 from their SPARV pipeline (III/283/aux), of which 105 also have {\it Gaia} $v_{\rm broad}$ values. We adopt the corresponding $\langle v$\,sin\,$i\rangle_{\rm ST}$
from the RAVE column in Table~\ref{table:Averagevsini} to calculate  $r_{\rm vsini}$. The LAMOST medium-resolution survey \citep[MRS;][]{Sun2021} includes 12 astrometric binaries in our final sample, of which 8 also have {\it Gaia} $v_{\rm broad}$ measurements. For $\langle v$\,sin\,$i\rangle_{\rm ST}$, we adopt the values in the Average column in Table~\ref{table:Averagevsini}. Finally, we query both the 3$^{\rm rd}$ data release of GALAH \citep{Buder2021} and the 16$^{\rm th}$ data release of APOGEE \citep{Jonsson2020}. The spectral fitting pipelines for both of these surveys are catered toward solar-type stars and therefore underestimate rotational velocities for rapidly spinning early-type stars. We exclude the GALAH and APOGEE measurements if either their own projected rotational velocity exceeds $v$\,sin\,$i$ $>$ 100~km~s$^{-1}$ or the {\it Gaia} broadening parameter is above $v_{\rm broad}$ $>$ 100~km~s$^{-1}$. In Table~A1, we list our 26 systems with reliable GALAH measurements, of which 20 also have {\it Gaia} $v_{\rm broad}$ values, and list our 24 systems with reliable APOGEE parameters, of which 18 have {\it Gaia} $v_{\rm broad}$ measurements. Similar to our LAMOST-MRS subset, we adopt $\langle v$\,sin\,$i\rangle_{\rm ST}$ from the Average column in Table~\ref{table:Averagevsini} for both the GALAH and APOGEE systems in order to compute $r_{\rm vsini}$. Of the 917 astrometric binaries in our final sample, 852 (93\%) have {\it Gaia} $r_{\rm vsini}$, 272 (30\%) have ground-based $r_{\rm vsini}$, and 207 have both.

\begin{figure}[t!]
\includegraphics[scale = 0.62]{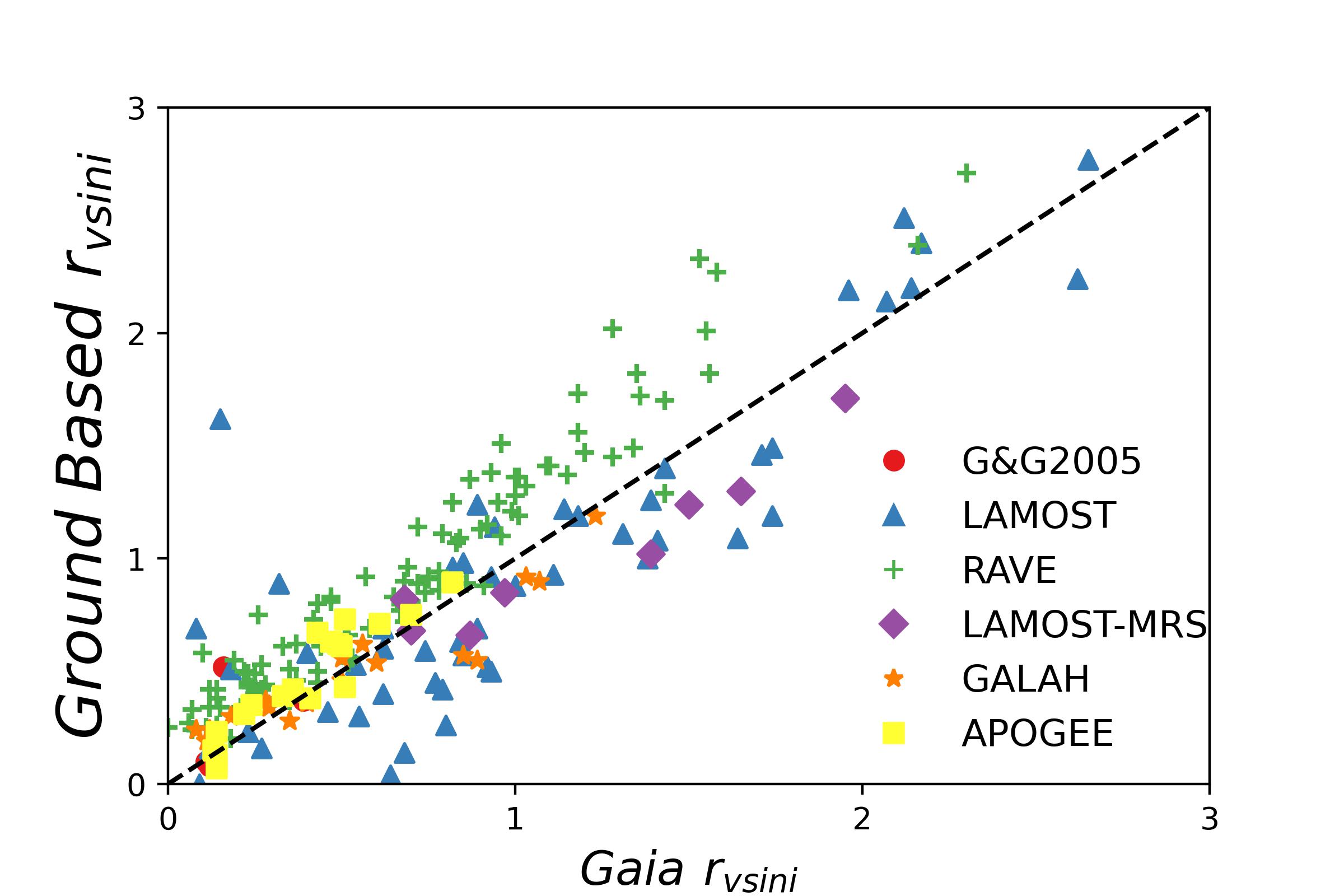}
\caption{The $r_{\rm vsini}$ measurements from various ground-based spectroscpopic surveys as a function of {\it Gaia} $r_{\rm vsini}$ for the 207 astrometric binaries common to both. Although some ground-based surveys exhibit small systematic offsets compared to {\it Gaia} (some in opposite directions from others), the overall systematic uncertainty in $r_{\rm vsini}$ is only 10\%\,-\,15\%. }
\label{fig:rcomp}
\vspace{3mm}
\end{figure}

\subsection{Comparison between Gaia and Ground-based $r_{\rm vsini}$}

\begin{figure*}[t!]
\begin{center}
\includegraphics[scale = 0.7]{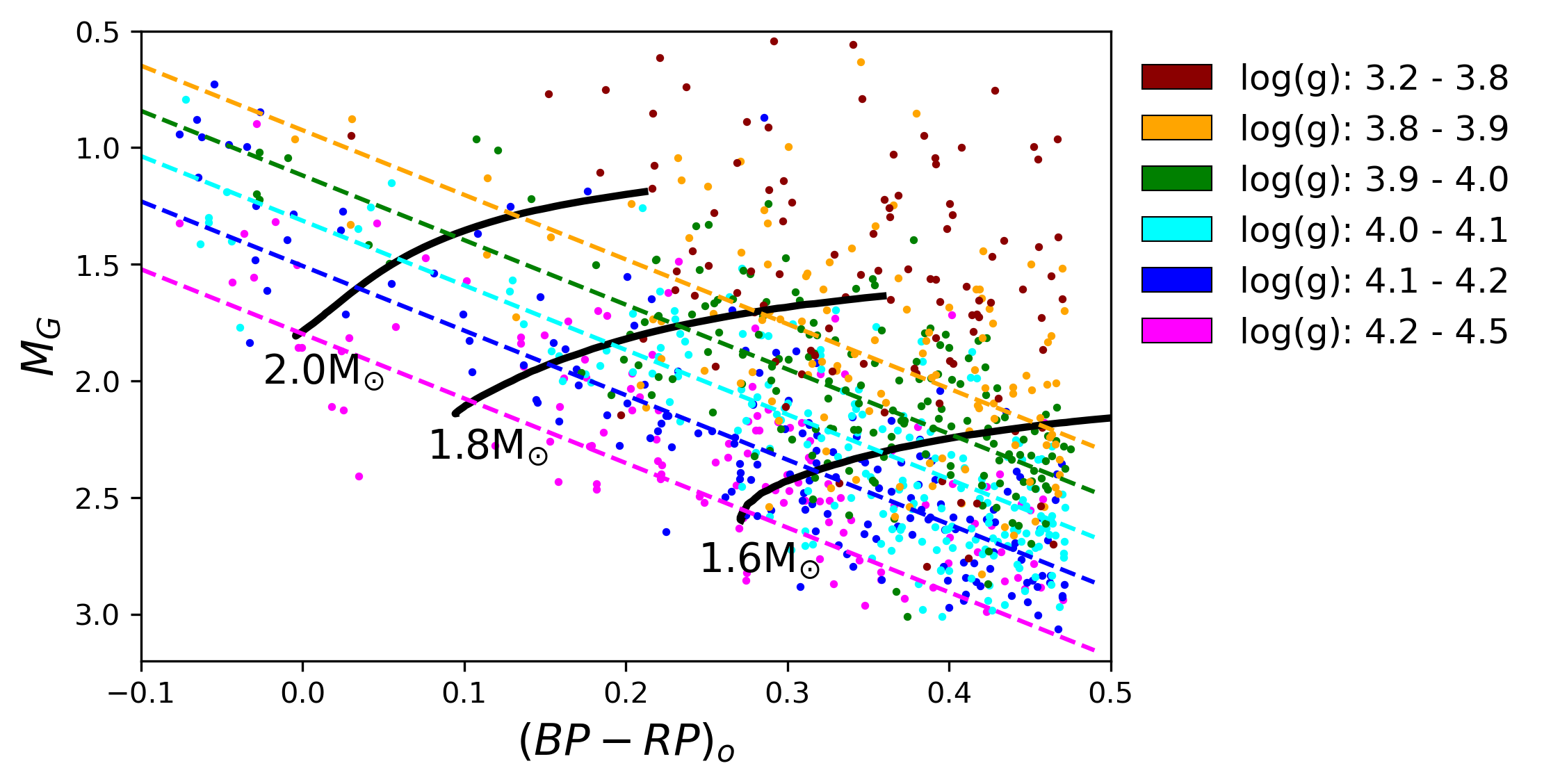}
\caption{Dust-corrected color-magnitude diagram of 884 astrometric binaries with spectroscopically measured $T_{\rm eff}$ and log\,$g$. We color our sample according to log\,$g$, including subgiants (log\,$g$ = 3.2\,-\,3.8; red) and five intervals for dwarfs (log\,$g$ = 3.8\,-\,4.5). We display our best-fit single-star relation for dwarfs (Eqn.~\ref{eqn:MGsingle}) at the midpoints of the corresponding log\,$g$ intervals (dashed). Solar-metallicity MIST evolutionary tracks for the middle of the MS (black) extend from log\,$g$ = 4.3 (bottom left) to 3.9 (top right) for the three specified masses.}
\label{fig:CMDmist}
\end{center}
\end{figure*}

To further demonstrate the accuracy of measuring projected rotational velocities from the {\it Gaia} broadening parameter, we compare $v$\,sin\,$i$ and $r_{\rm vsini}$ for the 207 systems common to both {\it Gaia} and ground-based spectroscopic surveys. We compute the differences in projected rotational velocity:

\begin{equation}
\Delta v\,{\rm sin}\,i = (v\,{\rm sin}\,i)_{\rm Gaia}~-~(v\,{\rm sin}\,i)_{\rm GroundBased}. 
\end{equation}

\noindent We report the mean offset $\mu$ (and standard deviation of the mean offset) and rms residuals $\sigma$ in Table~\ref{Tab:Gaia_vs_ground}. Similarly, we report the mean offsets and rms residuals for the differences in ratios $\Delta r_{\rm vsini}$. In Fig.~\ref{fig:rcomp}, we plot the individual ground-based $r_{\rm vsini}$ measurements as a function of {\it Gaia} $r_{\rm vsini}$ values.

The low-resolution LAMOST survey exhibits the largest scatter around the {\it Gaia} measurements, i.e., $\sigma$($\Delta r_{\rm vsini}$) = 0.36, but fortunately does not exhibit a systematic offset, i.e., $\mu$($\Delta$r$_{\rm vsini}$) = 0.05\,$\pm$\,0.05. The RAVE $v$\,sin\,$i$ and $r_{\rm vsini}$ measurements are consistently larger than the {\it Gaia} values by 17 km~s$^{-1}$ and 24\%, respectively. The APOGEE values are also systematically larger than {\it Gaia}, but to a lesser extent of 8 km~s$^{-1}$ and 7\%, respectively. Meanwhile, the LAMOST-MRS ratios $r_{\rm vsini}$ are consistently 18\% lower than the {\it Gaia} measurements. The GALAH velocities match the {\it Gaia} values quite well. 

After combining all 207 systems, the {\it Gaia} ratio $r_{\rm vsini}$ is on average 11\% smaller than the ground-based values, mostly driven by the large RAVE sample. \citet{Fremat2023} argued that the {\it Gaia} broadening parameter $v_{\rm broad}$ systematically underestimates $v$\,sin\,$i$ for systems fainter than $G$~$\gtrsim$~10. After limiting our comparison to the 37 astrometric binaries brighter than $G$~$<$~9.5, we indeed find a small reduction in the systematic offset, qualitatively consistent with \citet{Fremat2023}. However, the offset was reduced only slightly, from 11\% to 8\%. Instead of brightness effects, we surmise that the inherent differences between the various stellar atmospheric fitting pipelines are driving the small systematic offsets in $r_{\rm vsini}$, at least for our astrometric binary sample. Moreover, we already showed in section~\ref{sec:APRV} that systematic differences on the order of 5\%\,-\,20\% are already apparent when comparing the average $\langle v$\,sin\,$i \rangle_{\rm ST}$. By normalizing the individual {\it Gaia} $v$\,sin\,$i$ measurements, which comprise the majority of our sample, to their respective {\it Gaia} $\langle v$\,sin\,$i \rangle_{\rm ST}$ values in Table~\ref{table:Averagevsini}, we further minimize the systematic biases in $r_{\rm vsini}$. We thus conclude that the overall systematic uncertainty in our $r_{\rm vsini}$ measurements are only 10\%\,-\,15\% in our full sample of 917 astrometric binaries. For the 207 astrometric binaries that have both {\it Gaia} and ground-based 
{ spectra, we simply adopt their average $T_{\rm eff}$, log\,$g$, and $r_{\rm vsini}$ in our subsequent statistical analysis.}


\subsection{Basic Properties and Selection Biases}
\label{sec:basic}

As expected for our volume-limited sample of 917 astrometric binaries, the primaries are weighted toward later spectral types within our adopted interval. A substantial 403 (44\%) have F0-1\,IV/V primaries, 398 (43\%) are A5-9\,IV/V, 106 (12\%) are A0-4\,IV/V, and only 10 (1\%) have B8-9.5\,IV/V primaries. The average effective temperature is $T_{\rm eff}$ = 7,570~K, corresponding to an A8\,V primary with $M_1$ = 1.8\,\Msun\ \citep{Pecaut2013}.

In the following, we estimate dust reddenings $E$(BP-RP), dust-corrected absolute magnitudes $M_{\rm G}$, and primary masses $M_1$ for the 884 objects with spectroscopically measured $T_{\rm eff}$ and log\,$g$. We adopt the empirical relation in \citet{Pecaut2013} to map effective temperatures $T_{\rm eff}$ to intrinsic colors (BP-RP)${\rm o}$. We then compute the dust reddenings:

\begin{equation}
 E\mbox{(BP-RP)} = \mbox{(BP-RP)} - \mbox{(BP-RP)}_{\rm o}~. 
\end{equation}

\noindent  For the 35 systems with slightly negative dust reddenings $-$0.09~$<$~$E$(BP-RP)~$<$~0.00 based on this method, we set $E$(BP-RP)~=~0. The average dust reddening is only $\langle E$(BP-RP)$\rangle$ = 0.10 mag, as expected for our $d$ $<$ 500\,pc volume-limited sample. We adopt the standard {\it Gaia} dust extinction / reddening law appropriate for the Milky~Way \citep{Andrae2018}:

\begin{equation}
 A_{\rm G} = 2.0\,E\mbox{(BP-RP)}~.
\end{equation}

\noindent The absolute magnitudes are then:

\begin{equation}
M_{\rm G} = {\rm G}\,-\,5\,{\rm log}\,\Big(\frac{d}{\rm 10\,pc}\Big)\,-\,A_{\rm G}~.
\end{equation}

In Fig \ref{fig:CMDmist}, we present our dust-corrected color-magnitude diagram (CMD). We also distinguish systems according to their spectroscopically measured surface gravities. For a given (BP-RP)$_{\rm o}$, objects with lower log\,$g$ appear systematically brighter. We overlay solar-metallicity MIST evolutionary tracks \citep{Choi2016} across the truncated interval log\,$g$ = 3.9\,-\,4.3, which corresponds to the middle of the MS. Although the MIST evolutionary tracks follow key trends of the observed sample, there are systematic offsets due to the inherent differences among the various spectroscopic surveys and stellar fitting pipelines. For example, the ends of our displayed evolutionary tracks at log\,$g$ = 3.9 extend much further into the red than the data suggests.  We therefore rely on the \citet{Pecaut2013} empirical relation for dwarf stars to map effective temperature $T_{\rm eff}$ to primary masses $M_{\rm 1,PM13}$. We include a small correction term for surface gravity based on the computed MIST gradient between log\,$g$ and $M_1$ at fixed temperature:

\begin{equation}
M_1 (T_{\rm eff}, {\rm log}\,g) \approx M_{\rm 1,PM13}(T_{\rm eff}) - 0.9\,{\rm M}_{\odot}\,({\rm log}\,g - 4.1)~.
\end{equation}

\noindent The average primary mass of our sample is $\langle M_1 \rangle$ = 1.77 \Msun, similar to our estimate above based on spectral type. We report $A_{\rm G}$, $M_{\rm G}$, (BP$-$RP)$_{\rm o}$ and $M_1$ in Table~A1.

\begin{figure}[t!]
\includegraphics[scale = 0.6]{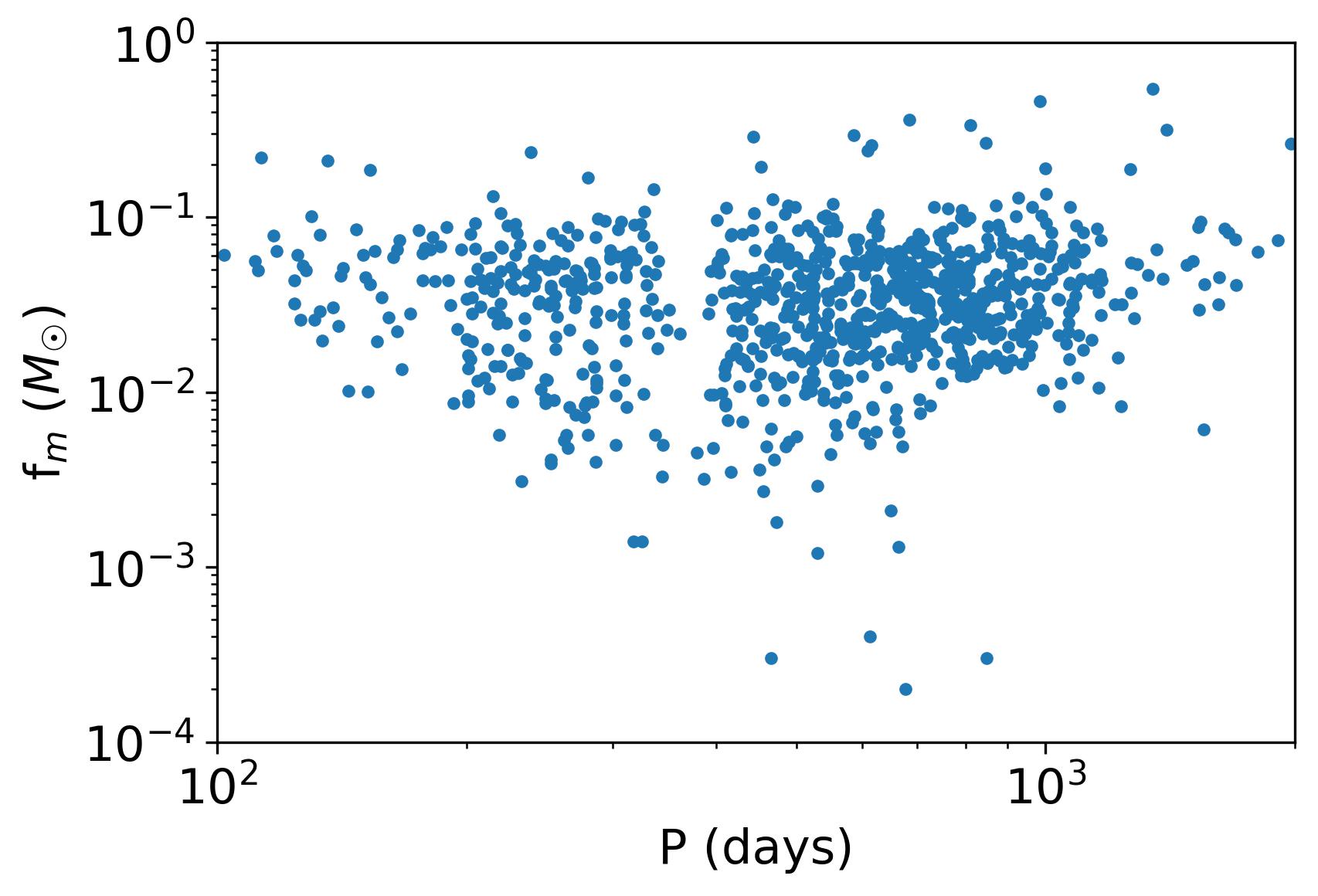}
\caption{Binary mass function $f_{\rm M}$ versus orbital period $P$ for our 917 astrometric binaries. Similar to the {\it Gaia} DR3 catalog as a whole, astrometric binaries near $P$ = 1~year are undetectable. At both very short and very long periods, only binaries with massive companions produce detectable and measurable astrometric orbits.}
\label{fig:fMvsP}
\vspace{2mm}
\end{figure}

Our astrometric binaries span orbital periods $P$~=~100\,-\,3,000~days, corresponding to $a$ = 0.5\,-\,5\,au. In Fig.~\ref{fig:fMvsP}, we compare the orbital periods and { astrometric binary} mass functions for our 917 astrometric binaries. As discussed in \citet{Halbwachs2023}, only the astrometric binaries with fitted orbital parameters above a certain quality threshold were reported in the {\it Gaia} DR3 catalog. Astrometric binaries with orbital periods close to one~year are impossible to distinguish from Earth's orbital motion, creating the observed gap at $P$~=~365~days in Fig.~\ref{fig:fMvsP} \citep[see also Fig.~3 in][]{Halbwachs2023}.

Extending toward short periods, only binaries with sufficiently large mass functions induce detectable astrometric shifts. Similarly, at longer periods, only astrometric binaries with large mass functions exceed the \citet{Halbwachs2023} quality threshold due to the finite time span and cadence of the {\it Gaia} observations. Given the photometric fluxes $F_1$ and $F_2$ of the primary (brighter) and secondary (fainter) components, respectively, the astrometric binary mass function is (Eqn.~14 in \citealt{Halbwachs2023}):

\begin{equation}
\label{eqn:fmFull}
f_{\rm M} = \frac{|F_1 M_2 - F_2 M_1|^3}{(F_1 + F_2)^3(M_1+M_2)^2}~. 
\end{equation}

\noindent In the limit where the secondary flux is negligible compared to the primary flux, then the binary mass function reduces to (Eqn.~15 in \citealt{Halbwachs2023}):

\begin{equation}
\label{eqn:fmApprox}
f_{\rm M}(F_2 \ll F_1) = \frac{M_1 q^3}{(1+q)^2}~. 
\end{equation}

\noindent A twin binary with equally bright components cannot be detected as an astrometric binary because its photocenter does not shift during its orbit. Nearly all of our astrometric binaries have mass functions below $f_{\rm M}$ $<$ 0.24\,\Msun. Of the 11 exceptions with $f_{\rm M}$ $>$ 0.24\,\Msun, 8 have evolved subgiant primaries that dominate the overall flux. Our astrometric binary sample becomes highly incomplete below $f_M$ $<$ 0.003\,\Msun (see Fig.~\ref{fig:fMvsP}), which corresponds to $q$~$>$~0.85 in the nearly twin limit (Eqn.~\ref{eqn:fmFull} and $F_2/F_1\,\approx\,(M_2/M_1$)$^{3.5}$) or $q$ $<$ 0.13 when the secondary flux is negligible (Eqn.~\ref{eqn:fmApprox}).


For our 100 astrometric binaries with luminous, subgiant primaries (log\,$g$ = 3.2\,-\,3.8), we estimate $q$ according to Eqn.~\ref{eqn:fmApprox}. For dwarf binaries (log\,$g$ = 3.8\,-\,4.5), companions with $q$ = 0.70\,-\,0.85 contribute a modest flux $F_2$ = 0.3\,-\,0.6\,$F_1$ and are still detectable via {\it Gaia} astrometry. An astrometric binary with a small mass function $f_{\rm M}$ $<$ 0.02\,\Msun\ can contain either a small $q$ $<$ 0.27 companion according to Eqn.~\ref{eqn:fmApprox} or a large $q$~$>$~0.70 companion according to Eqn.~\ref{eqn:fmFull}. To break this degeneracy, we measure the photometric excess $\Delta M_{\rm G}$ in search of photometric near-twins. We first divide our sample of 774 astrometric binaries with dwarf primaries into six intervals of (BP-RP)$_{\rm o}$ and five intervals of log\,$g$. For each (BP-RP)$_{\rm o}$\,-\,log\,$g$ grid point, we compute the 60$^{\rm th}$ percentile in the $M_G$ distribution (slightly fainter than median), which roughly corresponds to the absolute magnitude of a single star. We fit this single-star relation as a linear combination of color index and logarithmic surface gravity:

\begin{equation}
\label{eqn:MGsingle}
 M_{\rm G, single} \approx 1.41 + 2.77\, \rm {(BP}\mbox{-}\rm{RP)}_{\rm o} + 1.94\,({\rm log}\,g - 4.1)~.
\end{equation}

\noindent  We display in Fig.~\ref{fig:CMDmist} our fit to $M_{\rm G, single}$ as a continuous function of (BP-RP)$_{\rm o}$ and for our five intervals of log\,$g$.

The photometric excess is simply:

\begin{equation}
 \Delta M_{\rm G} = M_{\rm G, single} - M_{\rm G}~.
\end{equation}

\begin{figure}[t!]
\includegraphics[scale = 0.6]{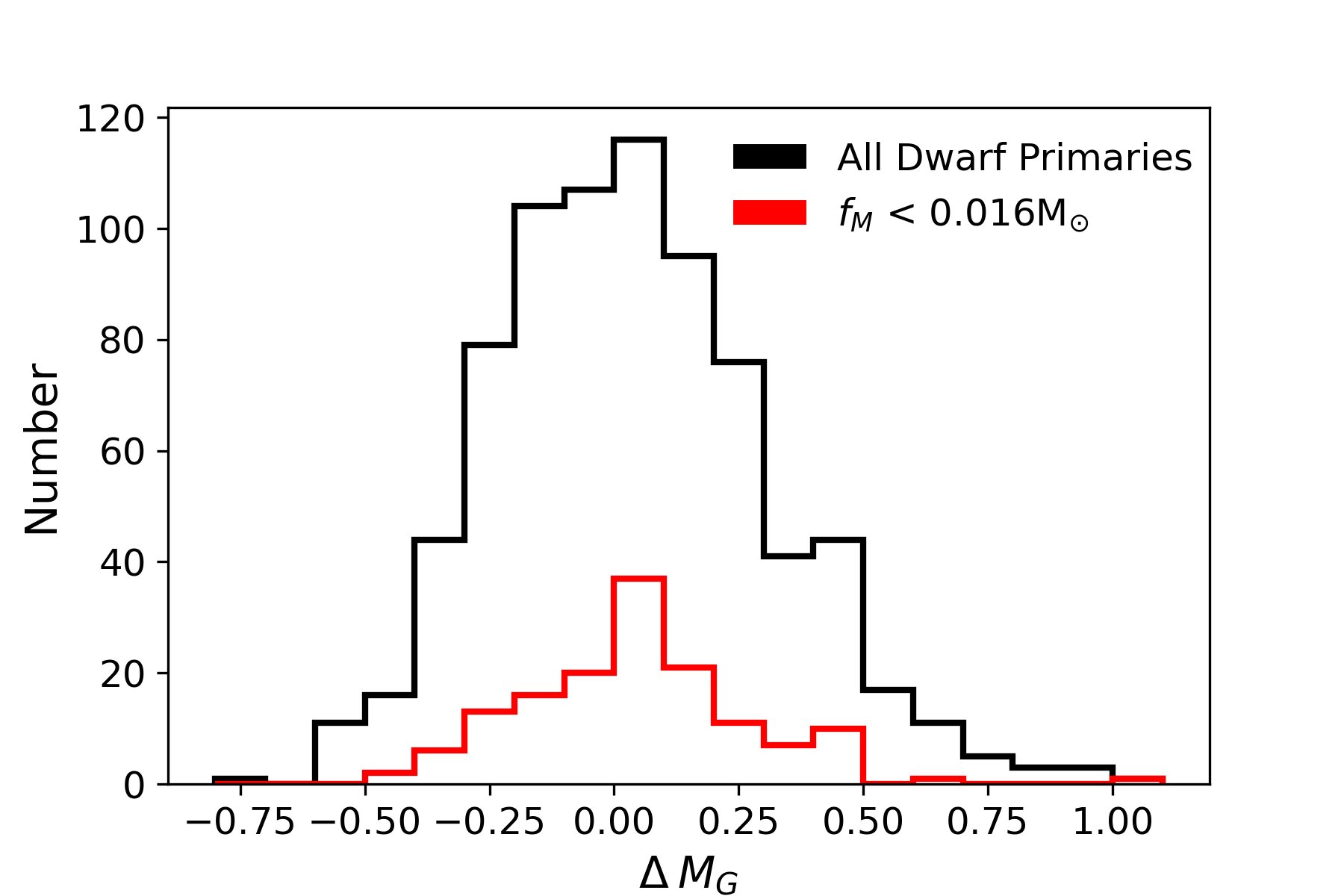}
\caption{The distribution of photometric excess $\Delta M_{\rm G}$ for our astrometric binaries with dwarf primaries (black) and the subset with $f_M$ $<$ 0.016\,\Msun\ (red). Both samples, especially the subset with small $f_{\rm M}$, exhibit a larger tail above $\Delta M_{\rm G}$ $>$ 0.35~mag, providing evidence for photometric near-twins with $q$ = 0.70\,-\,0.85.} 
\label{fig:delMGhist}
\vspace{4mm}
\end{figure}

\noindent  In Fig.~\ref{fig:delMGhist}, we display the distribution of $\Delta M_{\rm G}$ for our astrometric binaries with dwarf primaries. The distribution is slightly asymmetric with a longer tail extending toward positive $\Delta M_{\rm G}$, providing evidence for photometric near-twins. We also plot the $\Delta M_{\rm G}$ distribution for the subset with $f_M$ $<$ 0.016\,\Msun, which hints at a slightly bimodal distribution as expected for a population of binaries with mostly $q$ $<$ 0.24 and $q$ $>$ 0.73. A Kolmogorov-Smirnov (KS) test reveals that the $\Delta M_{\rm G}$ distributions for astrometric binaries with $f_M$ $<$ 0.016\,\Msun\, versus $f_M$ $>$ 0.016\,\Msun\ are discrepant at the $p_{\rm KS}$ = 0.013 (2.5\,$\sigma$) level, confirming that photometric near-twins prefer larger $\Delta M_{\rm G}$. Of our 784 astrometric binaries with dwarf primaries, only 101 (13\%) have substantive photometric excess $\Delta M_{\rm G}$ $>$ 0.35~mag. For a significant majority of our sample, the stellar parameters $T_{\rm eff}$, log\,$g$, and $v$\,sin\,$i$ are thus reliable measurements of the primaries because the secondaries are too faint to cause significant blending or systematic biases.

We divide our astrometric binary sample into three mass-ratio intervals based on $\Delta M_{\rm G}$ and estimates of $q$ from Eqn.~\ref{eqn:fmApprox}. We identify 372 (42\%) with $\Delta M_{\rm G}$ $<$ 0.35 and $q$ $<$ 0.32, 376 (43\%) with $\Delta M_{\rm G}$ $<$ 0.35 and $q$ = 0.32\,-\,0.50, and 136 (15\%) with either $\Delta M_{\rm G}$ $>$ 0.35 or $q$ $>$ 0.50 (mostly photometric near-twins). We report the corresponding mass-ratio bin for each of our binaries in Table~A1. Late-A/early-F binaries with intermediate separations are weighted toward small mass ratios such that only 24\% have $q$ = 0.5\,-\,0.85 \citep{Murphy2018}. {\it Gaia} is not sensitive to $q$ $>$ 0.85 twin binaries and is highly incomplete to $q$ = 0.6\,-\,0.85 near-twin binaries. It thus not surprising that only 15\% of our astrometric binary sample falls into the $q$ $>$ 0.5 bin.



\begin{figure}[t!]
\includegraphics[scale = 0.6]{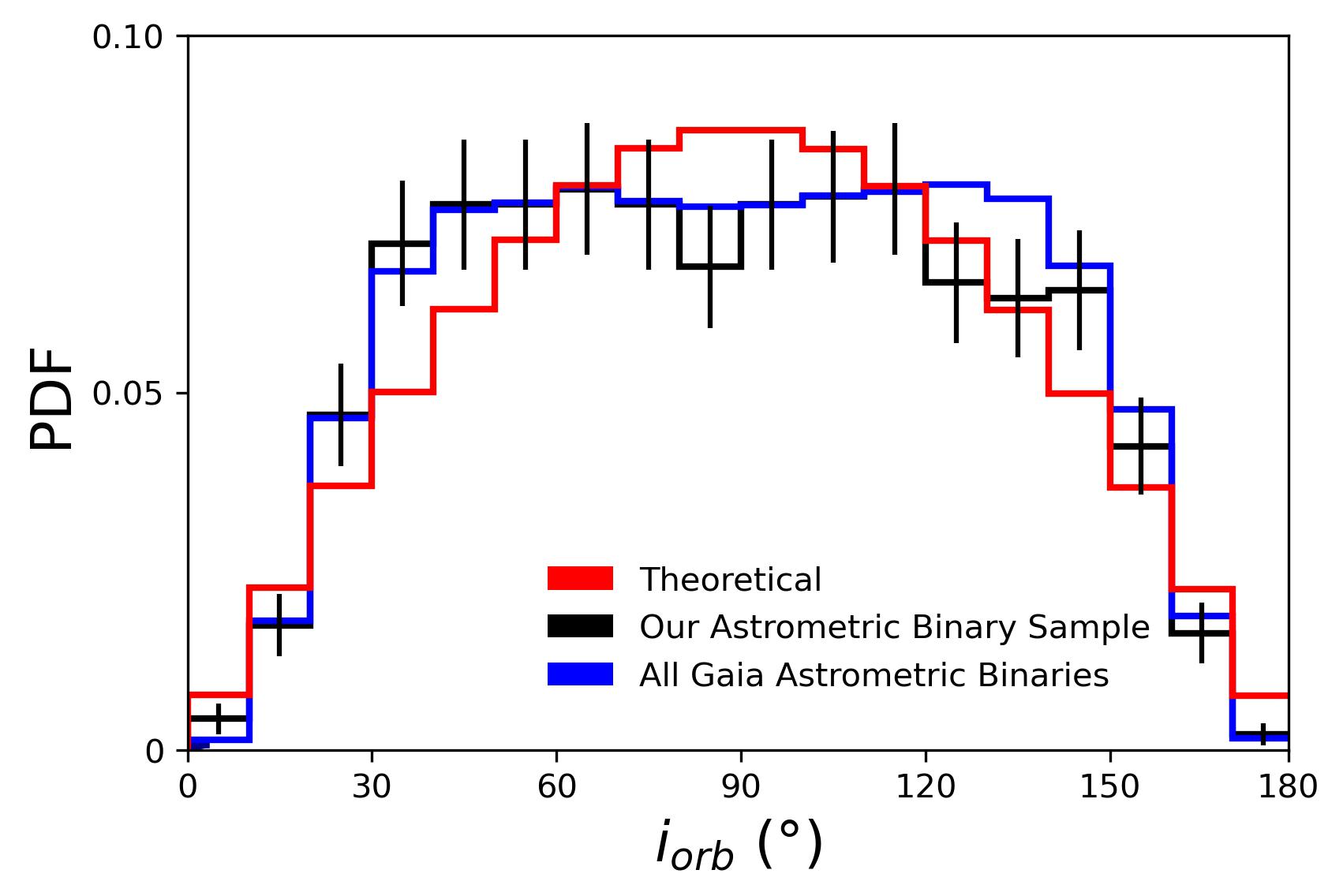}
\caption{The observed orbital inclination distribution of all {\it Gaia} astrometric binaries (blue) and our selected subset of 917 astrometric binaries (black) compared to the theoretical distribution expected from random orientations (red). Both nearly face-on and nearly edge-on astrometric binaries are slightly under-represented in {\it Gaia} DR3, which we account for in our statistical analysis.}
\label{fig:PDFinc}
\vspace{4mm}
\end{figure}

Given the finite cadence of the {\it Gaia} observations, astrometric binaries with face-on and edge-on orientations are slightly more difficult to characterize compared to those with moderate inclinations \citep{Halbwachs2023}. In Fig.~\ref{fig:PDFinc}, we compare the probability density functions (PDFs) of $i_{\rm orb}$ in 10$^{\circ}$ bins for all 165,500 {\it Gaia} astrometric binaries and our final subset of 917 systems. Using a Monte Carlo technique, we overlay the theoretical PDF expected from random orientations, whereby the inclinations are generated via $i_{\rm orb}$ = arccos($x$) where $x$ = U[$-$1,1] is a uniform random variable. Both the full {\it Gaia} astrometric binary sample and our subset exhibit similar distributions that deviate from theoretical expectations. A KS test reveals a discrepancy between our astrometric binary sample and random orientations that is statistically significant at the $p$~=~0.005 (2.8$\sigma$) level. Despite the selection bias in the {\it Gaia} DR3 astrometric binary catalog, we can still compare the overall $r_{\rm vsini}$ distributions as a function of $i_{\rm orb}$. In our modeling and fitting of the data (see section~\ref{sec:MC}), we simply must bootstrap the observed distribution of $i_{\rm orb}$ instead of assuming random orientations.

\subsection{Spin-Orbit Alignment}

We now compare the normalized ratios $r_{\rm vsini}$ of projected rotational velocity as a function of binary orbital inclination $i_{\rm orb}$. We perform three different statistical tests, and we present our results in Table~\ref{Tab:results} for the full sample and multiple subsets. We first compute the Spearman rank correlation coefficient $\rho$ between $r_{\rm vsini}$ and sin\,$i_{\rm orb}$, the corresponding probability $p_{\rho}$ of no correlation, and its statistical significance $\sigma_{\rho}$. We then divide the sample into three inclination bins: face-on orbits with sin\,$i_{\rm orb}$ = 0\,-\,0.5, i.e., $i_{\rm orb}$ = 0$^{\circ}$\,-\,30$^{\circ}$ or 150$^{\circ}$\,-\,180$^{\circ}$, middle systems with sin\,$i_{\rm orb}$ = 0.5\,-\,0.866, i.e., $i_{\rm orb}$ = 30$^{\circ}$\,-\,60$^{\circ}$ or 120$^{\circ}$\,-\,150$^{\circ}$, and edge-on orbits with sin\,$i_{\rm orb}$ = 0.866\,-\,1, i.e., $i_{\rm orb}$ = 60$^{\circ}$\,-\,120$^{\circ}$. We compute the average ratios $\langle r_{\rm vsini} \rangle$ for these three inclination bins and the population as a whole. We then calculate the difference $\Delta\langle r_{\rm vsini} \rangle$ between the average ratios of the face-on versus edge-on astrometric binaries and the statistical significance $\sigma_{\Delta}$ that it differs from zero. Finally, we perform a two-sample KS test of the cumulative distribution functions (CDFs) of $r_{\rm vsini}$ between our face-on and edge-on astrometric binaries. We report the resulting probability $p_{\rm KS}$ of no correlation and its statistical significance $\sigma_{\rm KS}$ in Table~\ref{Tab:results}.

\begin{figure}
\includegraphics[scale = 0.60]{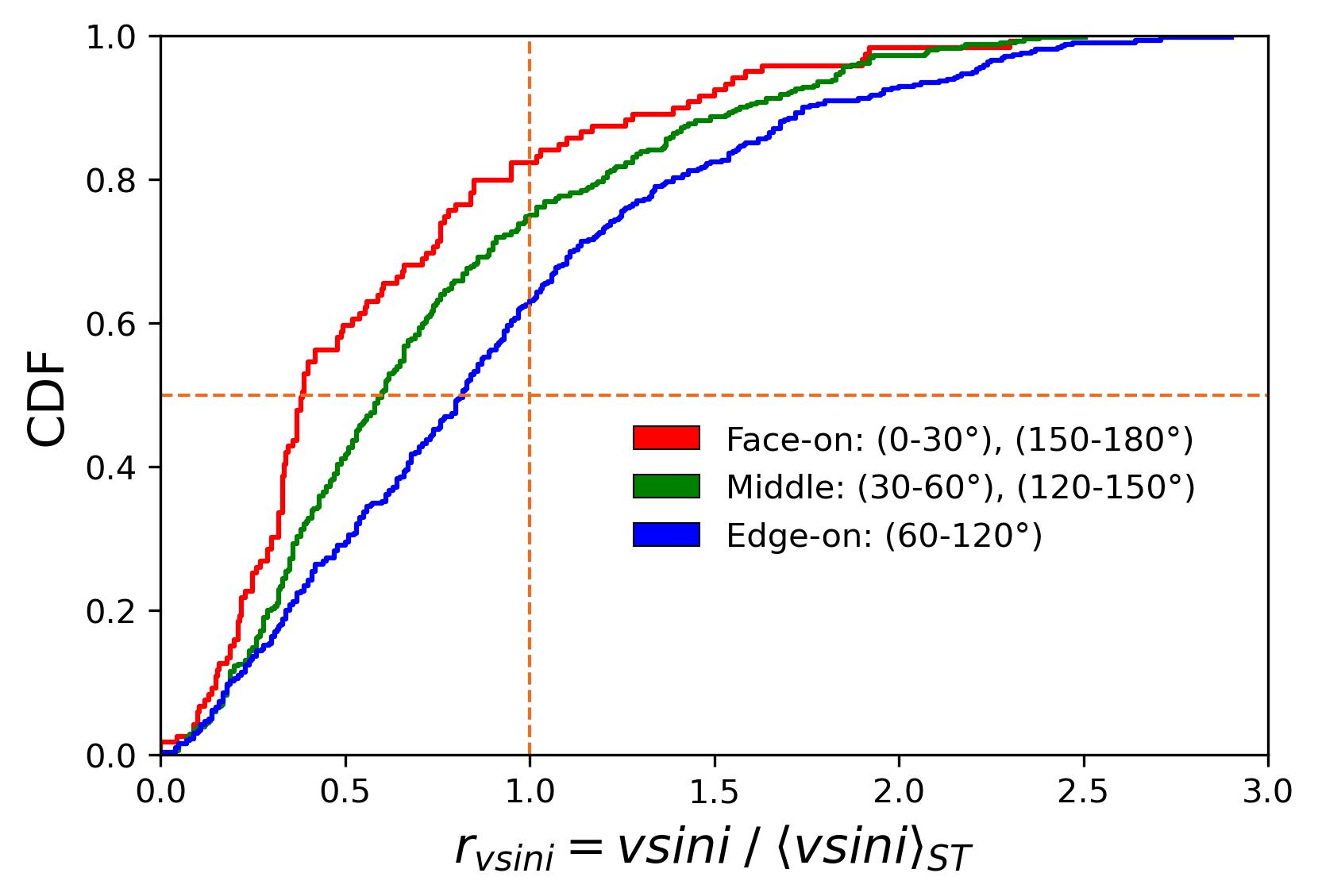}
\caption{Cumulative distribution functions of normalized projected rotational velocity ratios $r_{\rm vsini}$ for our full sample of 917 astrometric binaries separated into three orbital inclination bins: face-on (red), middle (green), and edge-on (blue). To guide the eye, $r_{\rm vsini}$ = 1 and CDF = 0.5 are plotted as orange dashed lines.  Astrometric binaries in face-on orbits contain primary stars that are skewed toward smaller $r_{\rm vsini}$ and thus favor pole-on spin orientations, demonstrating a substantial degree of spin-orbit alignment at the 6$\sigma$ significance level.} 
\label{fig:CDFfull}
\vspace*{3mm}
\end{figure}

\renewcommand{\arraystretch}{1.7}
\setlength{\tabcolsep}{2.5pt}
\begin{deluxetable*}{lccccccccccccccc}
\tabletypesize{\footnotesize}
\tablecaption{Projected Rotational Velocity Ratios $r_{\rm vsini}$ versus Orbital Inclinations $i_{\rm orb}$}
\startdata
~ & \multicolumn{5}{c}{All} & \multicolumn{2}{c}{Face-on~~~~~} & \multicolumn{2}{c}{Middle~~~~~} & \multicolumn{2}{c}{Edge-on~~~~~} & \multicolumn{4}{c}{Face-on versus Edge-on}
 \vspace*{-0.1cm} \\
 Sample & N & $\langle r_{\rm vsini}\rangle$ & $\rho$ & $p_{\rho}$ & $\sigma_{\rho}$ & N & $\langle r_{\rm vsini}\rangle$ & N & $\langle r_{\rm vsini}\rangle$ & N & $\langle r_{\rm vsini}\rangle$ & $\Delta \langle r_{\rm vsini}\rangle$ & $\sigma_{\Delta}$ & $p_{\rm KS}$ & $\sigma_{\rm KS}$ \\
 \hline
 Full & 917 & 0.80\,$\pm$\,0.02 & 0.23 & 4$\times$10$^{-12}$ & 6.8 & 124 & 0.58\,$\pm$\,0.04 & 390 & 0.74\,$\pm$\,0.03 & 403 & 0.92\,$\pm$\,0.03 & $-$0.33\,$\pm$\,0.05 & 6.1 & 2$\times$10$^{-9}$ & 5.9 \\
F0-1\,IV/V & 403 & 0.80\,$\pm$\,0.03 & 0.20 & 6$\times$10$^{-5}$ & 3.8 & 48 & 0.67\,$\pm$\,0.08 & 170 & 0.74\,$\pm$\,0.04 & 185 & 0.90\,$\pm$\,0.05 & $-$0.23\,$\pm$\,0.09 & 2.5 & 0.006 & 2.5  \\
A7-A9\,IV/V & 289 & 0.78\,$\pm$\,0.03 & 0.25 & 1$\times$10$^{-5}$ & 4.2 & 43 & 0.59\,$\pm$\,0.08 & 125 & 0.70\,$\pm$\,0.04 & 121 & 0.94\,$\pm$\,0.06 & $-$0.35\,$\pm$\,0.09 & 3.7 & 3$\times$10$^{-5}$  & 4.0 \\
 B8-A6\,IV/V & 225 & 0.80\,$\pm$\,0.04 & 0.26 & 9$\times$10$^{-5}$ & 3.7 & 33 & 0.46\,$\pm$\,0.06 & 95 & 0.80\,$\pm$\,0.06 & 97 & 0.92\,$\pm$\,0.06 & $-$0.46\,$\pm$\,0.09 & 5.3 & 8$\times$10$^{-6}$  & 4.3  \\
 $q$ $<$ 0.32 & 372 & 0.75\,$\pm$\,0.03 & 0.26 & 4$\times$10$^{-7}$ & 4.9 & 49 & 0.48\,$\pm$\,0.06 & 170 & 0.71\,$\pm$\,0.04 & 153 & 0.89\,$\pm$\,0.05 & $-$0.41\,$\pm$\,0.07 & 5.6 & 6$\times$10$^{-6}$ & 4.4 \\
$q$ = 0.32\,-\,0.50 & 376 & 0.85\,$\pm$\,0.03 & 0.18 & 2$\times$10$^{-5}$ & 4.1 & 51 & 0.69\,$\pm$\,0.07 & 155 & 0.78\,$\pm$\,0.05 & 170 & 0.96\,$\pm$\,0.05 & $-$0.27\,$\pm$\,0.09 & 3.1 & 0.003 & 2.7 \\
$e$ $<$ 0.2 & 345 & 0.67\,$\pm$\,0.02 & 0.39 & 3$\times$10$^{-14}$ & 7.5 & 46 & 0.36\,$\pm$\,0.04 & 159 & 0.62\,$\pm$\,0.03 & 140 & 0.83\,$\pm$\,0.04 & $-$0.47\,$\pm$\,0.06 & 8.1 & 2$\times$10$^{-10}$ & 6.2 \\
$e$ = 0.2\,-\,0.4 & 286 & 0.81\,$\pm$\,0.03 & 0.15 & 0.01 & 2.3 & 30 & 0.64\,$\pm$\,0.08 & 118 & 0.75\,$\pm$\,0.05 & 138 & 0.89\,$\pm$\,0.05 & $-$0.25\,$\pm$\,0.10 & 2.6 & 0.13 & 1.1 \\
$e$ $>$ 0.4 & 286 & 0.95\,$\pm$\,0.04 & 0.15 & 0.009 & 2.3 & 48 & 0.77\,$\pm$\,0.09 & 113 & 0.92\,$\pm$\,0.06 & 125 & 1.04\,$\pm$\,0.06 & $-$0.27\,$\pm$\,0.11 & 2.5 & 0.008 & 2.4 \\
 $P$\,=\,100\,-\,400\,d & 249 & 0.82\,$\pm$\,0.04 & 0.14 & 0.03 & 1.8 & 34 & 0.60\,$\pm$\,0.07 & 102 & 0.79\,$\pm$\,0.05 & 113 & 0.93\,$\pm$\,0.06 & $-$0.33\,$\pm$\,0.09 & 3.6 & 0.04 & 1.8 \\
  $P$\,=\,400\,-\,700\,d & 367 & 0.83\,$\pm$\,0.03 & 0.24 & 4$\times$10$^{-6}$ & 4.5 & 48 & 0.64\,$\pm$\,0.08 & 156 & 0.75\,$\pm$\,0.04 & 163 & 0.96\,$\pm$\,0.05 & $-$0.31\,$\pm$\,0.09 & 3.3 & 3\,$\times$\,10$^{-5}$ & 4.0 \\
$P$\,=\,700\,-\,3,000\,d & 301 & 0.74\,$\pm$\,0.03 & 0.27 & 2$\times$10$^{-6}$ & 4.7 & 42 & 0.50\,$\pm$\,0.07 & 132 & 0.71\,$\pm$\,0.04 & 127 & 0.85\,$\pm$\,0.05 & $-$0.35\,$\pm$\,0.09 & 3.8 & 0.0005 & 3.3 
\enddata
 \tablecomments{The first six columns list the number and average ratio $\langle r_{\rm vsini} \rangle$ for all astrometric binaries in the listed sample, the Spearman rank correlation coefficient $\rho$ between $r_{\rm vsini}$ and sin\,$i_{\rm orb}$, and the corresponding probability $p_{\rho}$ and statistical significance $\sigma_{\rho}$. The next six columns give the number and average ratio $\langle r_{\rm vsini} \rangle$ of astrometric binaries in face-on orbits ($i_{\rm orb}$ = 0$^{\circ}$\,-\,30$^{\circ}$ or 150$^{\circ}$\,-\,180$^{\circ}$), middle orbits (30$^{\circ}$\,-\,60$^{\circ}$ or 120$^{\circ}$\,-\,150$^{\circ}$) and edge-on orbits (60$^{\circ}$\,-\,120$^{\circ}$). The last four columns compare the face-on versus edge-on astrometric binaries by listing the difference in average ratios $\Delta\langle r_{\rm vsini} \rangle$ and the KS probability $p_{\rm KS}$, and both their corresponding levels of significance.}
 \label{Tab:results}
\end{deluxetable*}
\setlength{\tabcolsep}{6pt}
\renewcommand{\arraystretch}{1.0}

For our full sample, $r_{\rm sini}$ and sin\,$i_{\rm orb}$ are positively correlated with a Spearman coefficient $\rho$~=~0.23 that is statistically significant at the 6.8$\sigma_{\rho}$ level. In Fig.~\ref{fig:CDFfull}, we display the CDFs of $r_{\rm vsini}$ for our three inclination intervals. The face-on astrometric binaries favor smaller $r_{\rm sini}$ than the edge-on orbits with a difference of $\Delta\langle r_{\rm vsini} \rangle$ = $-$0.33\,$\pm$\,0.05 that is statistically significant at the 6.1$\sigma_{\Delta}$ level. Similarly, a KS test demonstrates that the face-on astrometric binaries are skewed toward smaller $r_{\rm sini}$ compared to the edge-on systems at the 5.9$\sigma_{\rm KS}$ level. The primaries in face-on orbits clearly have smaller $r_{\rm sini}$ and therefore smaller $v$\,sin\,$i$, demonstrating pole-on spin orientations and thus a substantial degree of spin-orbit alignment. All three tests yield a similar 6$\sigma$ level of statistical significance.  The high degree of spin-orbit alignment demonstrates that the majority of intermediate-mass binaries across $a$ = 0.5\,-\,5~au accreted from a circumbinary disk. As we illustrate in sections~\ref{sec:Dep}\,-\,\ref{sec:MC}, the observed trends of spin-orbit alignment with various orbital properties further suggest that the majority of close early-type binaries likely formed via disk fragmentation and thus were born with their spins already aligned to their orbits.

\section{Origin of Slow Rotators}
\label{sec:SlowRotators}

\subsection{Serendipitous Discovery}

Surprisingly, our astrometric binaries contain primaries that rotate more slowly on average than their corresponding single stars and wide binaries of the same spectral type. The primaries in our full sample have an average normalized ratio of $\langle r_{\rm vsini} \rangle$ = 0.80\,$\pm$ 0.02, which is less than the expected value of unity at the 8$\sigma$ confidence level. Even our primaries in edge-on astrometric binaries are rotating more slowly, $\langle r_{\rm vsini} \rangle$ = 0.92\,$\pm$ 0.03 $<$~1. We conclude that early-type primaries in binaries with separations $a$~=~0.5\,-\,5~au comprise the slow-rotator population in the observed bimodal rotational velocity distribution.

\subsection{Previous Observations and Interpretations}

As discussed in section~\ref{sec:APRV}, the slow-rotator fraction increases with stellar mass. For early-F stars, the slow-rotator component is almost indistinguishable from the main population \citep{Royer2007}. The slow-rotator fraction quickly increases from 5\% for late-A stars to 10\%\,-\,20\% for late-B/early-A stars \citep[see Table~4 in][]{Zorec2012} and then to 25\% for early-B stars \citep{Dufton2013}. \citet{Royer2007} hypothesized that the bimodality originates from two different types of interactions or accretion modes with the surrounding pre-MS cirumstellar disk. Alternatively, \citet{Zorec2012} argued that the rotational velocities evolve rapidly at certain points during the MS due to redistribution of angular momentum, and thus the observed bimodal rotational velocity distribution is a MS age effect. They also speculated that tidal braking in very close binaries may further contribute to the slow-rotator population.

However, \citet{Huang2010} ruled out the tidal braking scenario for slow-rotators. They performed a detailed analysis of B-type stars in both the field and clusters of known ages. Utilizing multi-epoch spectra, they removed single-lined spectroscopic binaries (SB1s) with large radial velocity (RV) shifts $\Delta$RV~$>$~13~km~s$^{-1}$ between different nights, corresponding to very close binaries below $P$ $\lesssim$ 14~days. Yet the bimodal rotational velocity distribution persisted. Similarly, \citet{RamirezAgudelo2015} showed that O~stars in SB1s with $\Delta$RV $>$ 20~km~s$^{-1}$ and O~stars with constant RVs exhibit a similar broad peak near $v$\,sin\,$i$ = 150~km~s$^{-1}$. Most recently, \citet{Bodensteiner2023} could not distinguish any significant differences between the rotational velocity distributions of mid-B stars in SB1s versus those with constant RVs.

\citet{Huang2010} also showed that the rotation rates of B-type stars evolve during the MS but concluded that the evolution is too gradual and small to create a bimodal velocity distribution. They instead argued that slow rotators must form with their small rotational velocities during the pre-MS. \citet{Huang2010} noted that the slow-rotator fraction increases with stellar mass similar to the observed increase in the close binary fraction. They speculated that accretion from a circumbinary disk may deposit more of its angular momentum into the binary orbit rather than stellar spins. 

\citet{Dufton2013} measured a significant bimodal rotational velocity distribution of early-B stars in the very young Tarantula Nebula. Similar to \citet{Huang2010}, they concluded that early-type stars must form with a sizeable population of slow rotators and that MS evolution can only slightly modify the mean velocity of the distribution, not the overall shape. \citet{Dufton2013} argued that substantial pre-MS magnetic braking in a subset of early-B stars might explain the origin of slow rotators, albeit they concluded that binary related effects remain a possibility. 

\citet{Pinzon2021} compiled the projected rotational velocities of pre-MS stars, both T~Tauri ($M_1$ $<$ 2\,\Msun) and Herbig Ae/Be ($M_1$ = 2\,-\,10\,\Msun). While most of their Herbig Ae/Be stars are fast rotators above $v$\,sin\,$i$ $>$ 50 km~s$^{-1}$, $\approx$\,20\% are below $v$\,sin\,$i$ $<$ 30 km~s$^{-1}$, confirming that slow rotators are initially born with their small rotational velocities. \citet{Pinzon2021} flagged known binaries (open symbols in their Fig.~4), including wide visual systems, short-period RV variables, and double-lined spectroscopic binaries (SB2s). Interestingly, more than half of their slow-rotator Herbig Ae/Be stars are known binaries whereas only $\approx$\,10\% of the fast-rotator population are flagged as such. They did not comment on whether this dichotomy is real or a selection bias, e.g., it is easier to detect RV variables and SB2s if the primaries have narrower absorption features and thus smaller $v$\,sin\,$i$.

Recent high-precision photometry of several young clusters have revealed a split of late-B/A MS stars on the color-magnitude diagram \citep{Milone2016,Li2017,Milone2018,Li2019,Wang2022}. By modeling the oblateness and latitude-dependent temperatures of spinning stars, these studies determined that the blue sequence contains slow rotators, $\lesssim$35\% of break-up, while the red sequence contains fast rotators, $\gtrsim$60\% of break-up. Follow-up spectroscopy confirmed that the blue sequence stars are slowly rotating, $\langle v$\,sin\,$i \rangle$ = 70 km s$^{-1}$ on average, while the red sequence contains mostly rapid rotators, $\langle v$\,sin\,$i \rangle$ = 200 km s$^{-1}$ \citep{Marino2018,Bastian2018}. \citet{Wang2022} showed that the slow-rotator blue sequence follows a flat mass function while the rapid-rotator red sequence is consistent with a Salpeter initial mass function; i.e., the slow-rotator fraction increases with stellar mass. Additional hypotheses for the slow-rotator blue sequence were proposed, including rotational locking with long-lived pre-MS circumstellar disks \citep{Bastian2020} and MS binary merger products \citep{Wang2022}.

\subsection{Our Interpretation}
\label{sec:Interpretation}

Of all the proposed scenarios for the slow-rotator population of early-type stars, our conclusion comes closest to the pre-MS binary hypothesis of \citet{Huang2010} whereby some of the angular momentum of the circumbinary disk is transferred into the orbit. In addition, some of the disk angular momentum is probably channeled into the spin of the secondary. Given the same total angular momentum $J_{\rm tot}$, the spin angular momentum of a single star, $J_1$ = $J_{\rm tot}$, must be larger then that of a primary in a binary, $J_1$ = $J_{\rm tot}$\,$-$\,$J_{\rm orb}$\,$-$\,$J_2$. The physical processes of pre-MS accretion and torques between stellar spins and disks are uncertain, but in the following we discuss some key differences between single stars and binaries.

\citet{ArmitageClarke1996} modeled magnetic braking and spin evolution for both single and binary T~Tauri stars. For single stars, they assumed disk radii of 20~au and an initial accretion rate of 10$^{-7}$\,\Msun\,yr$^{-1}$. At early times, the magnetospheric radius remained relatively constant near the corotation radius of the Keplerian disk, thereby regulating the stellar rotation period at a roughly constant $P_{\rm spin}$~=~8~days. After 10~Myr, the disk mass sufficiently decreased such that the magnetic fields expelled the inner disk material to well beyond the corotation radius. After the star-disk magnetic linkage was broken, the star continued to contract toward the zero-age MS while spinning up by a factor of a few to $P_{\rm spin}$ = 2\,-\,4 days, consistent with the observed rotation periods of weak-lined T~Tauri stars. \citet{ArmitageClarke1996} then simulated binary stars across different periastron separations $r_{\rm peri}$ = 1\,-\,8 au, and the circumprimary disk was tidally truncated to 0.4\,$r_{\rm peri}$. They assumed the same initial primary accretion rate of 10$^{-7}$\,\Msun\,yr$^{-1}$, and therefore the surface densities of the tidally truncated circumprimary disks were substantially larger in their binary models. The magnetic fields could no longer completely expel the inner disk at late times, and thus magnetic braking sustained a roughly constant $P_{\rm spin}$ = 8 days throughout their entire binary simulation. The primaries in all of their binary models across $r_{\rm peri}$ = 1\,-\,8\,au ended as slow rotators, whereby the spin angular momentum of the primary was transferred through the disk into the orbit of the secondary. For a slightly closer binary, \citet{ArmitageClarke1996} argued that the circumprimary disk would be sufficiently small that the magnetic fields could completely expel the circumprimary disk at late times, resulting in a rapid rotator. Meanwhile, the circumprimary disk in a slightly wider binary would mimic the disk profiles of single stars, and therefore the primary would also evolve into a rapid rotator. According to the \citet{ArmitageClarke1996} models, only binaries across intermediate separations $a$~$\approx$~0.5\,-\,10~au produce slow rotators, which is remarkably consistent with our astrometric binary population.

One major difference from the T Tauri stars modeled by \citet{ArmitageClarke1996} is that our early-type stars are expected to accrete their final disk material with fully radiative envelopes where magnetic braking may not be as efficient. The measured gas masses of Herbig Ae/Be disks can reach $\approx$1\% the stellar mass \citep{Boissier2011,Grant2023}, which if accreted at the Keplerian frequency of the star-disk boundary can spin up an intermediate-mass star from 20\% to 40\% of break-up. However, some young OB stars exhibit fossil remnants of 1\,-\,2 kG magnetic fields \citep{Donati2006,Wade2006,Alecian2008,Martins2010}, which may have contributed to magnetic braking during their formation process. \citet{Rosen2012} modeled magnetic braking and spin evolution of single stars across a broad range of masses. For their fiducial input parameters, which include a 2~kG magnetic field, an initial disk mass that is 2\% the final stellar mass, and an accretion timescale of 1~Myr, a 3\Msun\ star achieves 40\% critical rotation. By decreasing the initial disk mass to $<$\,0.5\% the final stellar mass or by accelerating the accretion timescale to $<$\,0.2~Myr, \citet{Rosen2012} demonstrated that a 3\,\Msun\ star will become a slow rotator below $<$\,20\% of break-up. In short, a B/A star can become a slow rotator if it accretes $\gtrsim$\,99.5\% of its final mass while still a large, cool, pre-MS star below the Kraft break where magnetic braking can effectively spin down the star.


Given the same final stellar mass, the accretion history of the primaries in our astrometric binaries are systematically different from single stars or those in wider binaries. Specifically, there are two reasons to suggest that early-type primaries in close binaries accreted a comparatively larger fraction of their final mass at earlier times while still below the Kraft break. First, the primaries that exhibit both spin reduction and spin-orbit alignment likely formed companions via disk fragmentation, which requires a massive protostellar disk and large accretion rates to drive the gravitational instability \citep{Kratter2008,Kratter&Lodato2016}. Close binaries that formed via disk fragmentation and inward disk migration represent the stochastic outliers that achieved sufficiently large accretion rates at early times. If instead the companion formed via core fragmentation and hardened to $a$~$<$~10~au after $>$\,0.2~Myr, i.e., after the early-type primary evolved above the Kraft break, then the primary would be a fast rotator like its single-star and wide-binary counterparts. 

Second, hydrodynamic models of circumbinary disks demonstrate that most of the mass and therefore angular momentum are accreted by the secondary \citep{Farris2014,Young2015}. As the low-mass companion sweeps out a larger area in its orbit and comes closer to the inner edge of the circumbinary disk, it preferentially accretes a substantially larger fraction of the infalling disk material. Specifically, \citet{Young2015} determined that for cold gas relative to the binary orbital velocity, which is appropriate for close binaries within $a$~$<$~10~au, the primary in a $q$ = 0.2 mass-ratio binary will accrete only $\approx$\,10\% of the gas from the circumbinary disk. As expected, this fraction increases to 50\% for twin binaries with $q$~=~1. As discussed in section~\ref{sec:basic}, our astrometric binaries are almost exclusively $q$~$<$~0.8 binaries where we naturally expect the secondaries to accrete most of the final disk mass and angular momentum.


In summary, disk fragmentation and inward disk migration require that the primary accreted comparatively more of its final mass at early times while still below the Kraft break. In this regime, the magnetic braking model of \citet{ArmitageClarke1996} can effectively spin down primaries in binaries across $a$ = 0.5\,-\,10 au by transferring angular momentum from the primary's spin through the disk and into the orbit. Even after the pre-MS primary contracts and evolves above the Kraft break, the low-mass companion accretes most of the remaining mass in the circumbinary disk \citep{Farris2014,Young2015}, efficiently directing angular momentum flow away from the primary and into the spin of the secondary. The combination of these two effects offer testable predictions for how the slow-rotator population varies with binary properties, which we presently analyze.

\section{Dependence on Binary Parameters}
\label{sec:Dep}

\subsection{Spectral Type}

We now investigate the degree of spin-orbit alignment and reduction in rotation rate as a function of binary parameters, and we report our results in Table~\ref{Tab:results}. We first divide our full sample into three primary spectral type bins: F0-1\,IV/V ($\langle M_1 \rangle$ = 1.5\,\Msun), A7-A9\,IV/V ($\langle M_1 \rangle$ = 1.8\,\Msun), and B8-A6\,IV/V ($\langle M_1 \rangle$ = 2.1\,\Msun). The average ratios of projected rotational velocities are all consistent with each other and match the $\langle r _{\rm vsini} \rangle$ = 0.80 value for the full sample. Hence the ratio in mean velocities between the slow-rotator versus fast-rotator populations does not vary significantly across $M_1$ = 1.5\,-\,3.0\,\Msun, which is consistent with observations \citep{Royer2007,Zorec2012,Wang2022}.

\begin{figure}
\includegraphics[scale = 0.62]{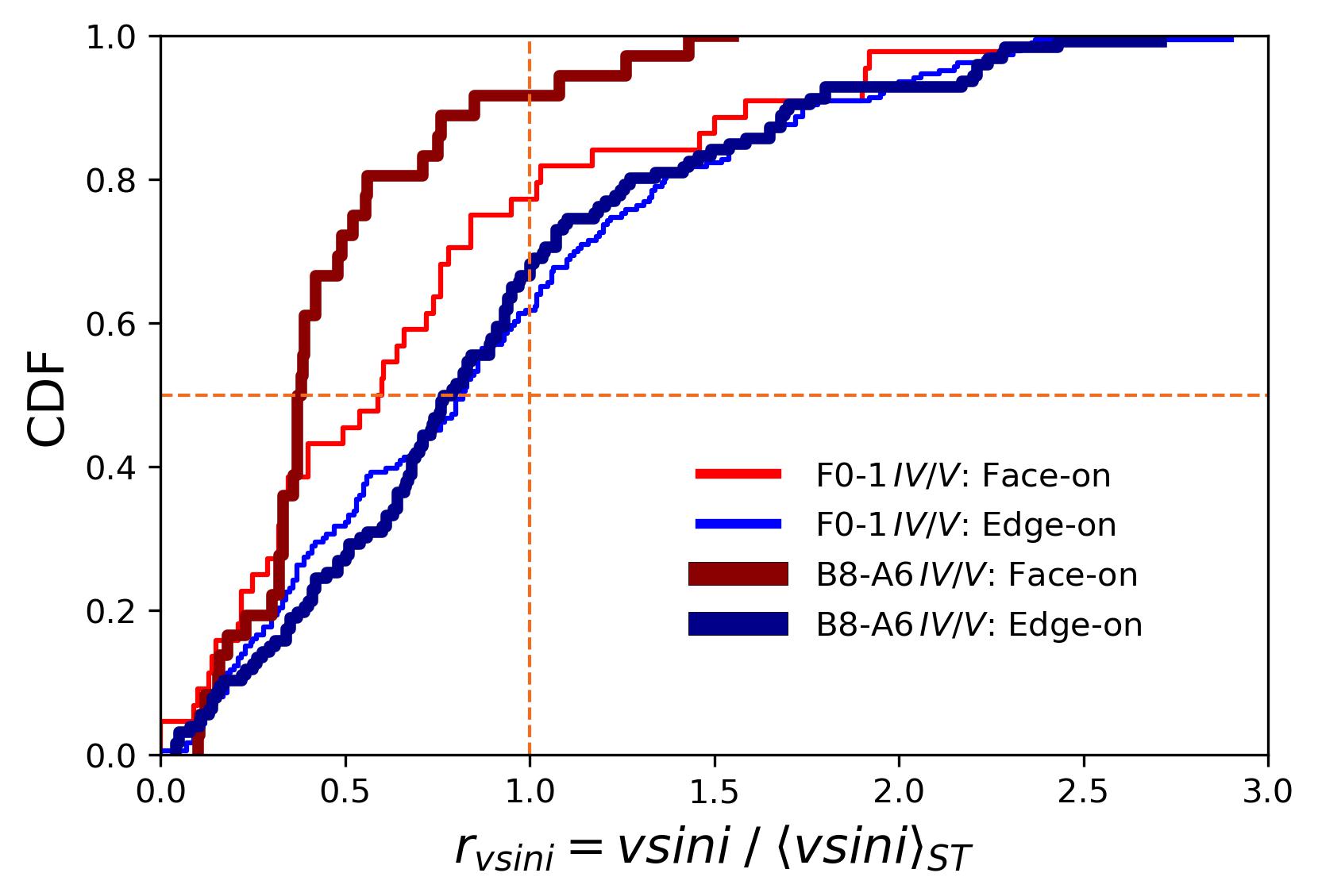}
\caption{Similar to Fig.~\ref{fig:CDFfull}, but for the subsets with B8-A6\,IV/V primaries (darker shades and thicker lines) and F0-1\,IV/V primaries (lighter shades and thinner lines) and only for the face-on (red) and edge-on (blue) astrometric binaries. } 
\label{fig:stcdf}
\vspace*{3mm}
\end{figure}

The B8-A6\,IV/V subset shows a greater degree of spin-orbit alignment compared to the F0-1\,IV/V subset. The correlation between $r_{\rm vsini}$ and sin\,$i_{\rm orb}$ is stronger for the B8-A6\,IV/V primaries ($\rho$ = 0.26) compared to the F0-1\,IV/V subset ($\rho$ = 0.20). Despite its smaller sample size, the B8-A6\,IV/V subset yields a larger statistical significance of spin-orbit alignment ($\sigma_{\Delta}$~=~5.3 and $\sigma_{\rm KS}$~=~4.3) compared to our F0-1\,IV/V sample ($\sigma_{\Delta}$ = $\sigma_{\rm KS}$ = 2.5).  We display in Fig.~\ref{fig:stcdf} the CDFs of $r_{\rm vsini}$ for both face-on and edge-on astrometric binaries in our F0-1\,IV/V and B8-A6\,IV/V subsets. For the face-on astrometric binaries with B8\,-\,A6 primaries (dark red CDF), a substantial 29/33 = 88\% have $r_{\rm vsini}$ $<$ 0.8.  

The fraction of astrometric binaries that have significant spin-orbit alignment must therefore increase with primary mass. More massive protostellar disks are more prone to gravitational instability and fragmentation \citep{Kratter2008}. The rapid rise in the close binary fraction $F_{\rm close}$ $\propto$ $M_1^{0.5}$ above $M_1$~$>$~1~\Msun\ has been interpreted as the increase in the propensity for disk fragmentation and inward disk migration \citep{Tokovinin2020,Offner2023}. The predicted trend between disk fragmentation and stellar mass is consistent with the observed increase in spin-orbit alignment across $M_1$ = 1.5\,-\,3.0\,\Msun.

\subsection{Mass Ratio}

\begin{figure}
\includegraphics[scale = 0.62]{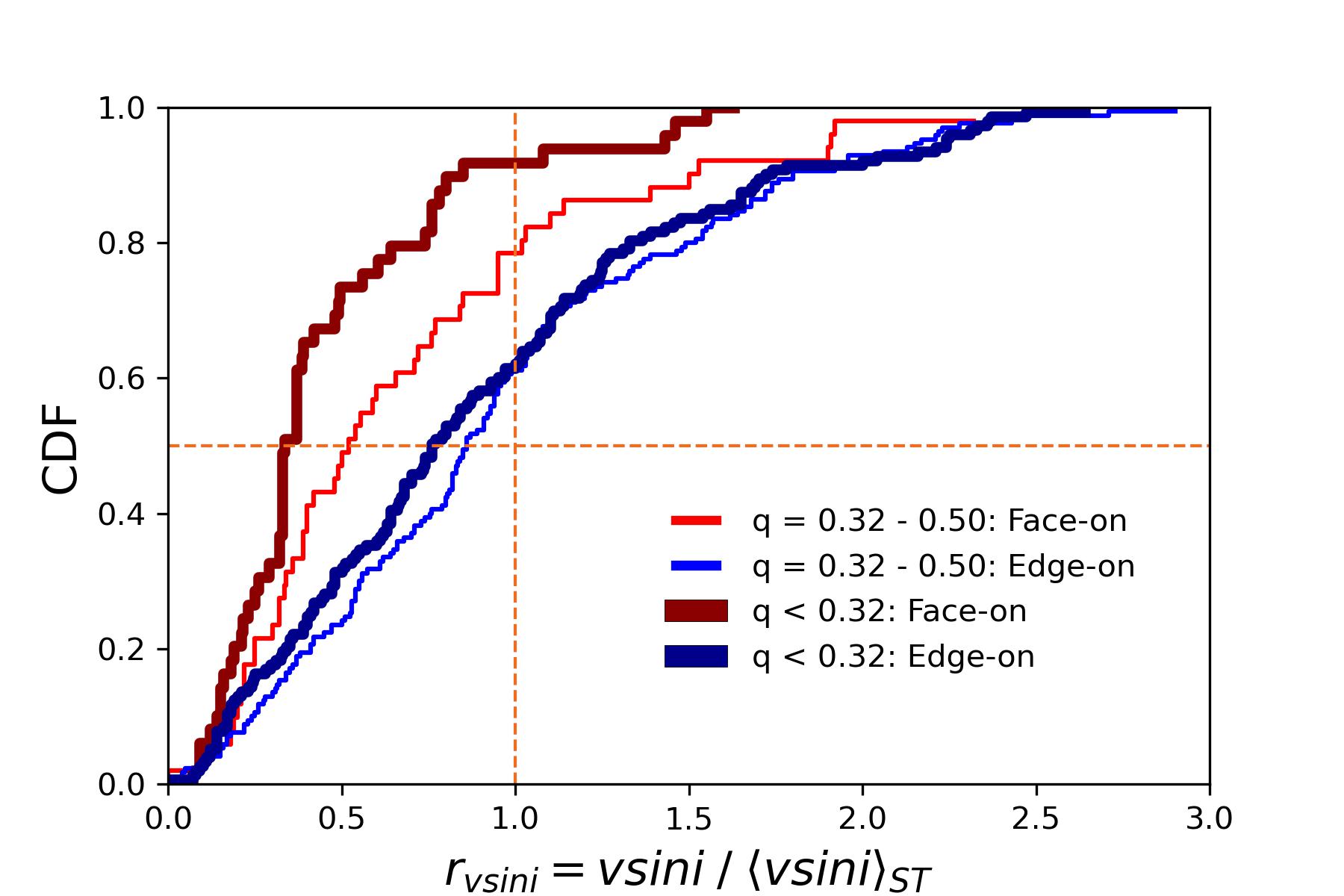}
\caption{Similar to Fig.~\ref{fig:CDFfull}, but for the subsets with binary mass ratios q $=$ 0.32-0.50 (lighter shades and thinner lines) and  q $<$ 0.32 (darker shades and thicker lines) and only for the face-on (red) and edge-on (blue) astrometric binaries.}
\label{fig:qCDF}
\vspace*{4mm}
\end{figure}

We next compare in Fig.~\ref{fig:qCDF} the degree of spin-orbit alignment for our astrometric binaries with small ($q$ $<$ 0.32) versus moderate ($q$ = 0.32\,-\,0.50) mass ratios. Although both samples are comparable in size, the small mass-ratio subset exhibits slower overall spins ($\langle r _{\rm vsini} \rangle$ = 0.75) and a much stronger degree of spin-orbit alignment ($\rho$ = 0.26, $\sigma_{\rho}$ = 4.9, $\sigma_{\Delta}$ = 5.6, and $\sigma_{\rm KS}$ = 4.4). For the face-on astrometric binaries with $q$ $<$ 0.32 (dark red CDF), a significant majority 43/49 = 88\% have $r_{\rm vsini}$ $<$ 0.8. Meanwhile, astrometric binaries with moderate mass ratios $q$ = 0.32\,-\,0.50 have faster spins ($\langle r _{\rm vsini} \rangle$ = 0.85) but are still rotating more slowly than their single-star counterparts. Similarly, our subset with moderate mass ratios exhibits a weaker but nonetheless statistically significant degree of spin-orbit alignment ($\rho$ = 0.18, $\sigma_{\rho}$ = 4.1, $\sigma_{\Delta}$ = 3.1, and $\sigma_{\rm KS}$ = 2.7).


As discussed in section~\ref{sec:Intro}, disk fragmentation and inward disk migration can more readily produce close extreme mass-ratio binaries with primary spins that are already aligned to their orbits. Moreover, the observed trend with { mass ratio} is consistent with our interpretation in section~\ref{sec:Interpretation} that low-mass secondaries accrete a larger fraction of the circumbinary disk mass and angular momentum. For example, according to the \citet{Young2015} models, a { primary in a $q$ = 0.5 binary accretes 25\% of the circumbinary disk mass} while the primary in a $q$ = 0.1 binary accretes only 5\%. Thus the primaries in extreme mass-ratio astrometric binaries have the slowest spins within the slow-rotator population. 

We do not compute the statistics for our small sample of 136 astrometric binaries with $q$~$>$~0.50 due to competing selection biases. The degree of spin-orbit alignment may be exaggerated for photometric near-twins due to blending of their respective absorption features. For example, a spectrum of a near-twin, edge-on astrometric binary taken near quadrature will have a larger $v$\,sin\,$i$ due to the opposite radial velocities of its components. However, the transition from photometric near-twins to non-twins is not definitive (see Fig.~\ref{fig:delMGhist}). Thus some small mass-ratio binaries with large errors in their measured stellar parameters may have leaked into our photometric excess subset with $\Delta M_{\rm G}$ $>$ 0.35 mag. We can only surmise that the trend continues whereby near-twin and twin binaries still have significant spin-orbit alignment but with faster overall spins.

\subsection{Eccentricity}
\label{sec:ecc}

\begin{figure}
\includegraphics[scale = 0.61]{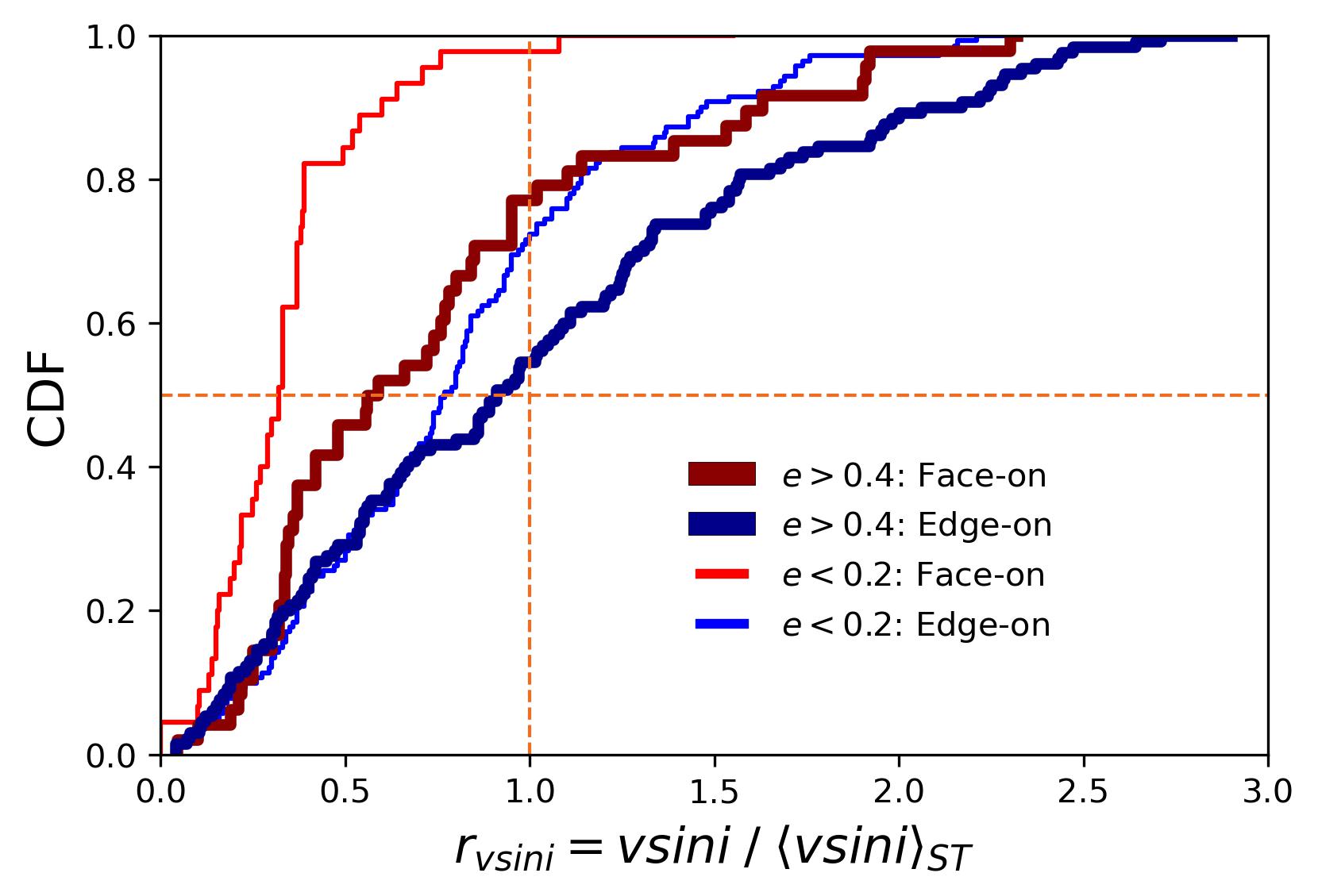}
\caption{Similar to Fig.~\ref{fig:CDFfull}, but for the subsets with eccentricities $e$ $<$ 0.2 (lighter shades and thinner lines) and $e$ $>$ 0.4 (darker shades and thicker lines) and only for the face-on (red) and edge-on (blue) astrometric binaries. } 
\label{fig:ecentcdf}
\vspace*{2mm}
\end{figure}

We next separate our full sample into three eccentricity bins: $e$ $<$ 0.2, $e$ = 0.2\,-\,0.4, and $e$ $>$ 0.4 (see Table~\ref{Tab:results}). In Fig.~\ref{fig:ecentcdf}, we display the CDFs of $r_{\rm vsini}$ for face-on and edge-on astrometric binaries in our $e$ $<$ 0.2 and $e$ $>$ 0.4 subsets. Our astrometric binaries with $e$ $<$ 0.2 exhibit a substantial reduction in spin ($\langle r _{\rm vsini} \rangle$ = 0.67) and much stronger spin-orbit alignment ($\rho$ = 0.39, $\sigma_{\rho}$ = 7.5, $\sigma_{\Delta}$ = 8.1, and $\sigma_{\rm KS}$ = 6.2). For the face-on astrometric binaries in our $e$ $<$ 0.2 subset (light red CDF in Fig.~\ref{Tab:results}), a substantial 44/46 = 96\% have $r_{\rm vsini}$ $<$ 0.8 and even 36/46 = 78\% have $r_{\rm vsini}$ $<$ 0.4. 

Circumbinary disk accretion tends to dampen the eccentricity toward circular orbits \citep{Bate2010,Bate2012,Bate2018}. However, as discussed in section~\ref{sec:Intro}, the protostellar disks of early-type primaries are extremely short lived. Core-fragmentation companions that dynamically migrate to short separations around early-type primaries likely achieve, on average, only minor reduction in eccentricity and spin-orbit angles (see also below). The small subset of core-fragmentation binaries that achieve the necessary circumbinary disk accretion to dampen the eccentricities to below $e$ $<$ 0.2 will also evolve toward twin mass ratios, which is contrary to our measurement above that close binaries with extreme mass ratios exhibit the strongest degree of spin-orbit alignment. It is thus more plausible that the majority of close early-type binaries formed via disk fragmentation and inward disk migration, which can produce systems with simultaneously small eccentricities, small mass ratios, and aligned, reduced primary spins.

Meanwhile, our astrometric binaries with $e$~$>$~0.4 exhibit neither reduction in spin ($\langle r _{\rm vsini} \rangle$ = 0.95\,$\pm$\,0.04) nor spin-orbit alignment ($\rho$ = 0.15, $\sigma_{\rho}$ $\approx$ $\sigma_{\Delta}$ $\approx$ $\sigma_{\rm KS}$ = 2.4). Such binaries likely formed via core fragmentation on large scales followed by orbital decay toward shorter separations via dynamical friction, resulting in moderately eccentric orbits. The observed correlation between spin-orbit alignment and eccentricity is consistent with the \citet{Bate2012} simulation results discussed in section~\ref{sec:Intro}. Considering our 631 astrometric binaries with $e$ $<$ 0.4 exhibit measurable reduction in spin, we can already estimate the fraction $F_{\rm align}$ = 631/917 $\approx$ 70\% that have reduced, aligned spins, most likely due to forming through disk fragmentation and circumbinary disk accretion.

\subsection{Orbital Period}

\begin{figure}
\includegraphics[scale = 0.61]{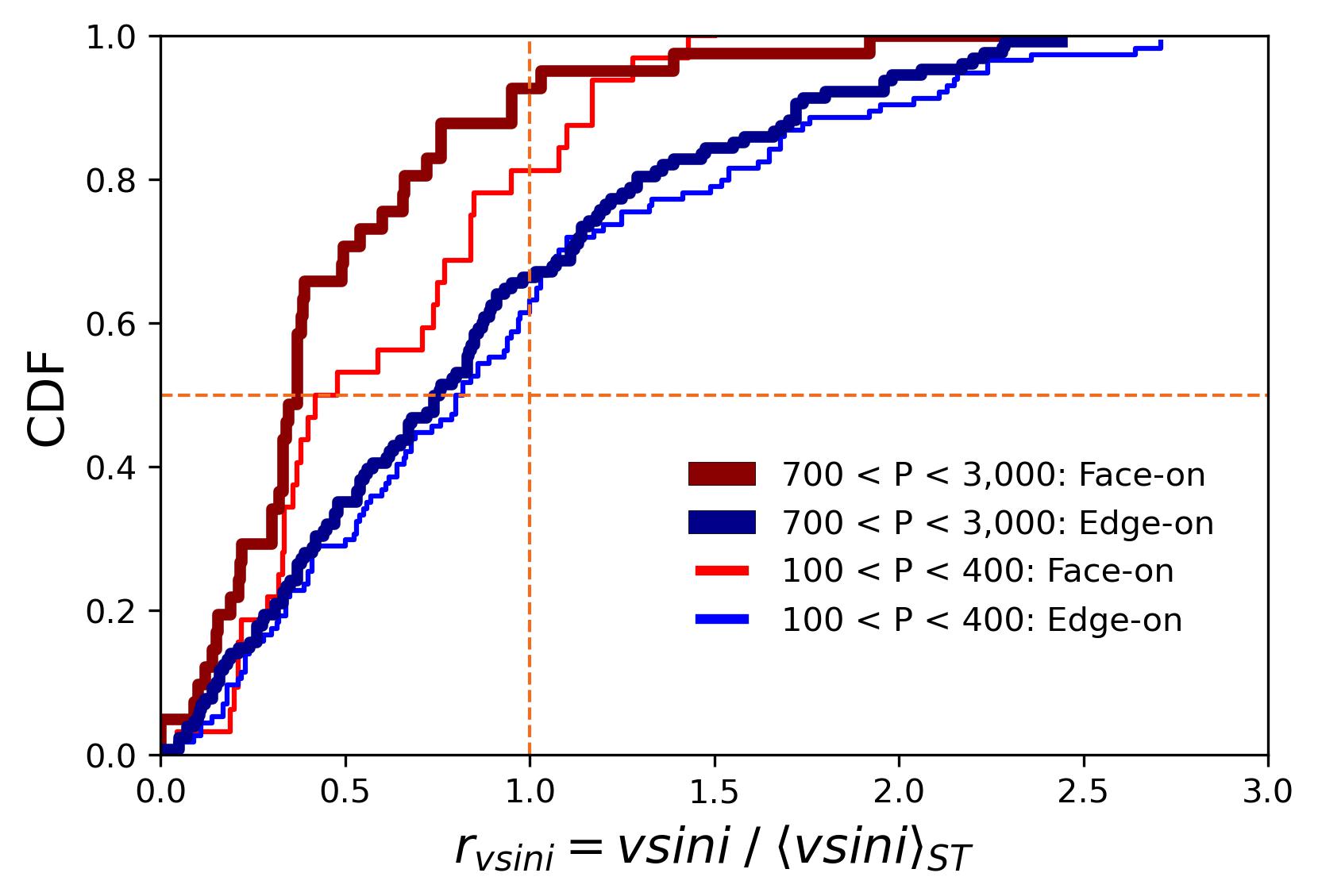}
\caption{Similar to Fig.~\ref{fig:CDFfull}, but for the subsets with orbital periods $P$ = 100\,-\,400 days (lighter shades and thinner lines) and $P$ $=$ 700\,-\,3,000~days (darker shades and thicker lines) and only for the face-on (red) and edge-on (blue) astrometric binaries.} 
\label{fig:Pcdf}
\vspace*{2mm}
\end{figure}

We finally divide our sample into three period intervals: $P$ = 100\,-\,400 days, $P$ = 400\,-\,700 days, and $P$ = 700\,-\,3,000 days (see Table~\ref{Tab:results}). In Fig.~\ref{fig:Pcdf}, we display the CDFs of $r_{\rm vsini}$ for face-on and edge-on astrometric binaries in our short-period and long-period subsets. Interestingly, the astrometric binaries across $P$ = 100\,-\,400~days ($a$ = 0.5\,-\,1.3~au) exhibit a slightly weaker reduction in spin ($r_{\rm vsini}$ = 0.82) and spin-orbit alignment ($\rho$ = 0.14, $\sigma_{\rho}$ = 1.8, $\sigma_{\Delta}$ = 3.6, and $\sigma_{\rm KS}$ = 1.8) compared to the wider systems across $P$ = 700\,-\,3,000~days ($a$ = 1.9\,-\,5~au; $\rho$ = 0.27, $\sigma_{\rho}$ = 4.7, $\sigma_{\Delta}$ = 3.8, and $\sigma_{\rm KS}$ = 3.3). For the face-on astrometric binaries across $P$ = 700\,-\,3,000 days (dark red CDF in Fig.~\ref{fig:Pcdf}), 36/42 = 86\% have $r_{\rm vsini}$ $<$ 0.8. 

As discussed in \citet{ArmitageClarke1996} and our section~\ref{sec:Interpretation}, magnetic braking in binaries is effective only across a narrow interval of separations $a$ = 0.5\,-\,10 au. At closer separations, the circumprimary disk masses are sufficiently tidally truncated that magnetic braking cannot sufficiently spin down the primary. Indeed, previous multi-epoch spectroscopic observations have already demonstrated that RV variables, i.e., very close binaries with $P$ $<$ 15 days ($a$ $<$ 0.2~au), do not exhibit an excess of slow-rotators \citep{Huang2010,RamirezAgudelo2015,Bodensteiner2023}. We can thus map the slow-rotator population as a function of binary separation: strongest across $a$ = 2\,-\,5~au, slightly weaker across $a$ = 0.5\,-\,1~au, and negligible below $a$ $<$ 0.2~au.


\begin{figure*}[th!]
\begin{center}
\includegraphics[scale = 0.8]{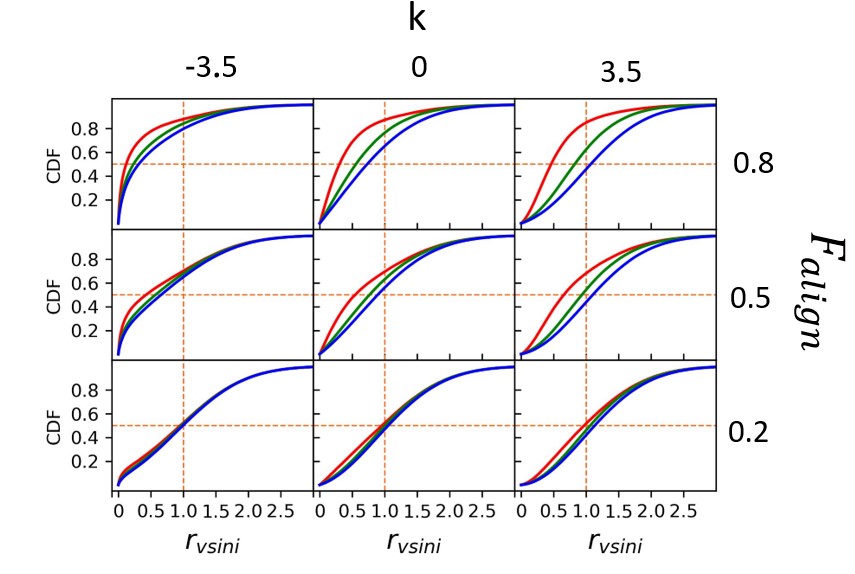}
\vspace*{-7mm}
\caption{Theoretical cumulative distributions of $r_{\rm vsini}$ for 3$\times$3 = 9 combinations of $F_{\rm align}$ and $k$, each corresponding to a simulated population of 10$^{5}$ binaries separated into the same three orbital inclination intervals: face-on (red), middle (green), and edge-on (blue). As the fraction $F_{\rm align}$ of binaries that exhibit spin-orbit alignment increases, the face-on and edge-on distributions diverge.  A small value of $k$ = $-$3.5 corresponds to substantial reduction in spin, $\langle S_{\rm align} \rangle$ = 0.25 on average, skewing the CDFs toward small ratios $r_{\rm vsini}$. Meanwhile, $k$ = 3.5 maps to $\langle S_{\rm align} \rangle$ = 0.75, only a minor reduction in spin.} 
\label{fig:CDF9x9}
\end{center}
\end{figure*}

\subsection{Occurrence Rate of Slow Rotators}

We have determined that primaries in binaries with $a$ = 0.5\,-\,10 au, $q$ $<$ 0.8, and $e$ $<$ 0.4 comprise the slow-rotator population in the bimodal rotational velocity distribution of early-type stars. We summarize and justify this narrow parameter space as follows. At closer separations, the circumprimary disk masses are sufficiently truncated that magnetic braking is ineffective \citep{ArmitageClarke1996}. Moreover, spectroscopic observations have already demonstrated that very close binaries within $a$ $<$ 0.2~au do not exhibit an excess of slow rotators \citep{Huang2010,RamirezAgudelo2015,Bodensteiner2023}. At wider separations, the circumprimary disk profiles and accretion evolution mimic those of single stars. In addition, B/A-type visual binaries beyond $a$ $>$ 30~au display neither spin-orbit alignment nor reduction in spin \citep{Slettebak1963,Levato1974,Weis1974}.  Disk fragmentation can more readily produce low-mass companions \citep{Kratter&Lodato2016}, which can accrete a larger fraction of mass and angular momentum from the circumbinary disk \citep{Farris2014,Young2015}, quenching the flow of angular momentum to the primary. The primaries in astrometric binaries with $q$ $<$ 0.32 have slower spins than those with $q$ = 0.32\,-\,0.50. Finally, eccentric astrometric binaries above $e$ $>$ 0.4 display neither spin reduction nor spin-orbit alignment.

The fraction of primaries that have companions across our narrow parameter space ($a$ = 0.5\,-\,10~au, $q$~$<$~0.8, and $e$~$<$~0.4) increases with primary mass according to \citep{Moe2017}:

\begin{equation}
 F = 0.06\,(M_1/{\rm M}_{\odot})^{0.6}\,.
\end{equation}

\noindent This relation yields 8\% for late-A/early-F primaries near $M_1$ = 1.6\,\Msun, 12\% for late-B primaries near $M_1$ = 3\,\Msun, and 24\% for early-B binaries near
$M_1$ = 10\,\Msun. The binary fraction across this parameter space closely matches the observed slow-rotator fraction as a function of primary mass \citep{Royer2007,Dufton2013}, which further corroborates our interpretations.

\section{Monte Carlo Modeling}
\label{sec:MC}

\subsection{Model Parameters and Population Synthesis}

We now fit parameterized models to the observed distributions of $r_{\rm vsini}$ in order to quantify the degree of spin-orbit alignment and spin reduction. Utilizing a Monte Carlo technique, we generate a population of binary stars based on two input model parameters. We first denote the fraction $F_{\rm align}$ of binaries that have primary spins aligned to the binary orbits, while the remaining fraction 1\,$-$\,$F_{\rm align}$ have random spin-orbit orientations. We then account for the slower rotation rates of the primaries in our aligned binaries with our second model parameter $k$. For the same fraction $F_{\rm align}$, we consider a distribution of spin reduction factors 0~$\le$~$S_{\rm align}$~$\le$~1 according to the coefficient $k$ in an exponential probability distribution:

\begin{equation}
p(S_{\rm align}) \propto {\rm exp}(k\,S_{\rm align}).
\label{eqn:exp}
\end{equation}

\noindent For example, $k$ = 0 corresponds to a uniform distribution of $S_{\rm align}$=U[0,1] with an average of $\langle S_{\rm align} \rangle$ = 0.5. Negative and positive values of $k$ result in distributions of $S_{\rm align}$ skewed toward smaller and larger values, respectively. Our analysis in section~\ref{sec:Dep} suggests that spin reductions follow a broad distribution that depends on a complex interplay of spectral type and orbital parameters $P$, $q$, and $e$. Moreover, we find that a distribution of $S_{\rm align}$ better matches the data compared to a singular value. For the remaining fraction 1\,$-$\,$F_{\rm align}$ of misaligned binaries, we simply assume no reduction in spin, i.e., $S_{\rm align}$ = 1.

For each combination of $F_{\rm align}$ and $k$, we synthesize 10$^{5}$ binaries. We select orbital inclinations $i_{\rm orb}$ according to random orientations, i.e., we draw cos\,$i_{\rm orb}$ = U[0,1] from a uniform distribution. When we actually fit models to the data (see section~\ref{sec:fitting}), we account for the small selection bias in $i_{\rm orb}$ within the Gaia DR3 astrometric binary catalog. We draw rotational velocities $v_{\rm rot}$ from a Maxwellian distribution across $v_{\rm rot}$ = 0\,-\,600~km~s$^{-1}$ that peaks at $v_{\rm rot}$ = 140 km s$^{-1}$, consistent with the { deprojected} velocity distribution of late-A/early-F stars that dominate our astrometric binary sample { \citep[][see Section~\ref{sec:APRV}]{Royer2007,Zorec2012}}. We eventually compute the normalized ratio $r_{\rm vsini}$ based on the weighted average $\langle v$\,sin\,$i \rangle_{\rm ST}$ of primary spectral types in our sample (see below), and thus there is no need to draw rotational velocities from mass-dependent distributions. 

For the fraction 1\,$-$\,$F_{\rm align}$ of binaries that are not aligned, we assume random spin orientations and select $i_{\rm spin}$ accordingly, fully independent of $i_{\rm orb}$. For these systems, the projected rotational velocity is simply $v$\,sin\,$i$ = $v_{\rm rot}$\,sin\,$i_{\rm spin}$. For the fraction $F_{\rm align}$ that is aligned, we select a spin reduction factor $S_{\rm align}$ for each system according to the exponential probability distribution given by $k$ in Eqn.~\ref{eqn:exp}. For simplicity, we assume perfect spin-orbit alignment, i.e., $i_{\rm spin}$ = $i_{\rm orb}$, and so the resulting projected rotational velocity is $v$\,sin\,$i$ = $S_{\rm align} v_{\rm rot}$\,sin\,$i_{\rm orb}$. By weighting the average projected rotational velocities $\langle v$\,sin\,$i \rangle_{\rm ST}$ in section~\ref{sec:APRV} with respect to the observed spectral types and luminosity classes within our full sample, we measure an overall average value of $\langle v$\,sin\,$i \rangle_{\rm ST}$ = 103~km~s$^{-1}$, similar to the projected rotational velocity of an A8\,V star. We finally compute $r_{\rm vsini}$ = $v$\,sin\,$i$\,/\,103\,km\,s$^{-1}$ for each binary accordingly.

\begin{figure}[t]
\includegraphics[scale = 0.61]{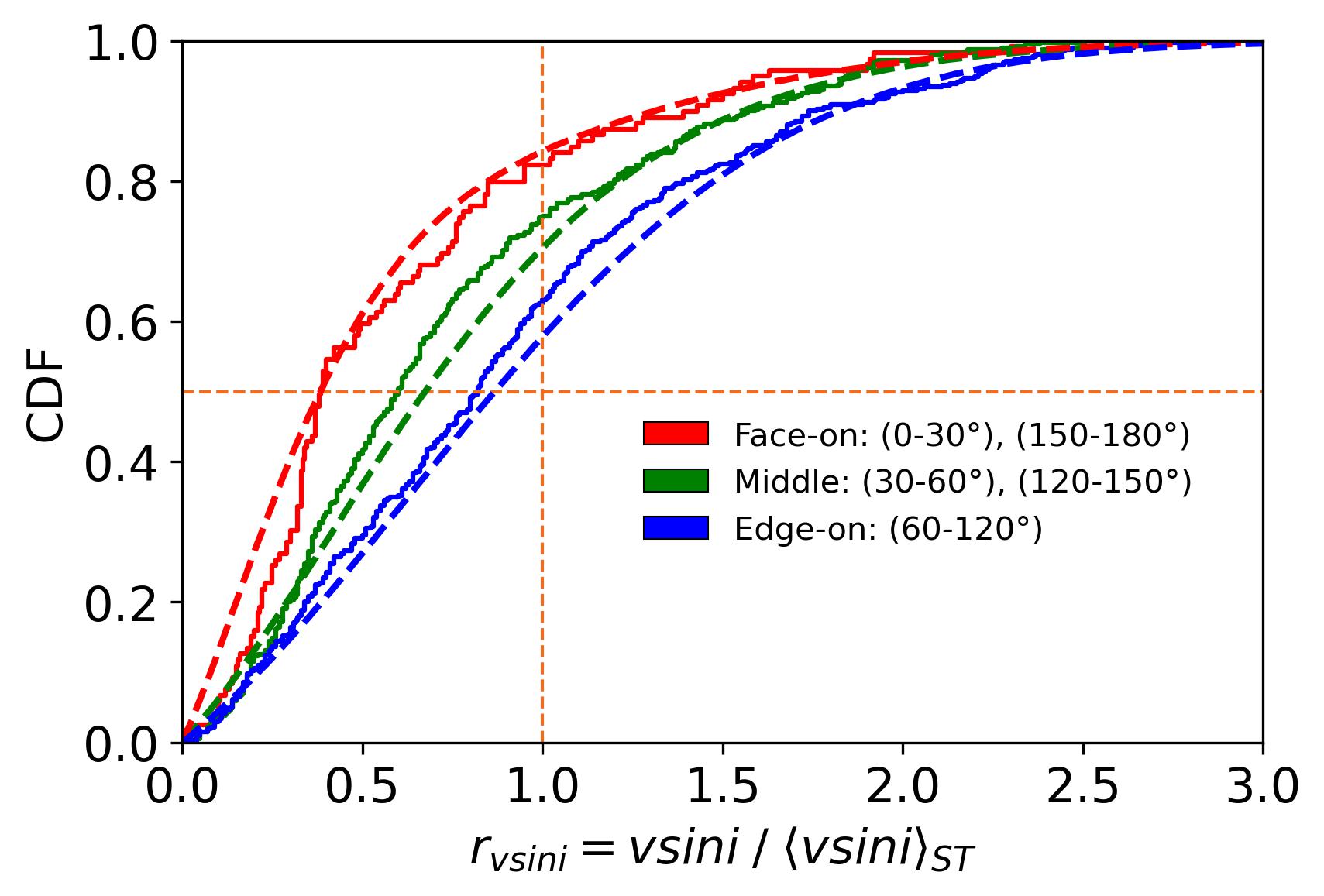}
\caption{Same data and format as Fig.~\ref{fig:CDFfull} (solid lines) but now overlaid with our best-fit Monte Carlo model (dashed lines): a spin-orbit alignment fraction of $F_{\rm align}$ = 0.76 and a coefficient of $k$ = $-$1.0, corresponding to an average spin reduction factor of $\langle S_{\rm align} \rangle$ = 0.42.}
\label{fig:CDFbfit}
\vspace*{6mm}
\end{figure}

We generate a large grid of models spanning 0~$\le$~$F_{\rm align}$~$\le$~1 and $-$3.5~$\le$~$k$~$\le$~3.5. In Fig~\ref{fig:CDF9x9}, we plot CDFs of $r_{\rm vsini}$ for 3$\times$3 = 9 combinations of $F_{\rm align}$ and $k$. Similar to Fig.~\ref{fig:CDFfull}, we distinguish three orbital inclination intervals, where face-on orbits with $i_{\rm orb}$~=~0\deg\,-\,30\deg\ or 150\deg\,-\,180\deg\ are shown in red, middle systems with $i_{\rm orb}$~=~30\deg\,-\,60\deg\ or 120\deg\,-\,150\deg\  are green, and edge-on orbits with $i_{\rm orb}$~=~60\deg\,-\,120\deg\ are blue. In all of our models with non-zero $F_{\rm align}$, the face-on orbits have smaller $r_{\rm vsini}$ as expected. With increasing $F_{\rm align}$ (bottom to top), the $r_{\rm vsini}$ cumulative distributions gradually diverge. For the left set of panels with $k$ = $-$3.5, the spin reduction factors according to Eqn.~\ref{eqn:exp} are weighted toward small values, i.e., $\langle S_{\rm align} \rangle$ = 0.25, skewing the resulting CDFs toward small $r_{\rm vsini}$. Meanwhile, the right set of panels with $k$ = 3.5 have $\langle S_{\rm align} \rangle$ = 0.75 and exhibit only modest spin reduction. By comparing to the observed distributions in Fig.~\ref{fig:CDFfull}, we can already discern by eye that the best-fit solution is somewhere close to $F_{\rm align}$ = 0.8 and between $k$ = $-$3.5 and 0.


\subsection{Fitting the Data}
\label{sec:fitting}

We adopt a maximum likelihood estimation method to fit our two model parameters $F_{\rm align}$ and $k$. We first separate our full sample of astrometric binaries into $N$~=~18 bins of orbital inclination $i_{\rm orb}$, each with 10\deg\ widths as displayed in Fig.~\ref{fig:PDFinc}. For each of the $M_i$ observed astrometric binaries within the $i^{\rm th}$ inclination bin, we calculate the probability $p_j(r_{{\rm vsini},j}(i_{{\rm orb},i})|F_{\rm align},k)$ for a given Monte Carlo model with parameters $F_{\rm align}$ and $k$. We compute the log-likelihood:

\begin{equation}
 {\rm ln}~L_i (i_{{\rm orb},i}|F_{\rm align},k) = \Sigma_{j=1}^{M_i}~{\rm ln}\,p_j(r_{{\rm vsini}_j}(i_{{\rm orb},i})|F_{\rm align},k)
\end{equation}

\noindent by summing over all observed $M_i$ astrometric binaries within that $i^{\rm th}$ inclination bin. In this manner, we weight our theoretical models that assumed random orbital orientations to the observed distributions of astrometric binary orbital inclinations, thereby accounting for the small selection bias discussed in section~\ref{sec:basic} and Fig.~\ref{fig:PDFinc}. We finally compute the overall log-likelihood function:

\begin{equation}
{\rm ln}~\mathcal{L} (F_{\rm align},k) = \Sigma_{i=1}^{N} {\rm ln}~L_i (i_{{\rm orb},i}|F_{\rm align},k)
\end{equation}

\noindent by summing over all $N$ = 18 inclination bins. We maximize ln~$\mathcal{L}$ to determine our best-fit solution for $F_{\rm align}$ and $k$. We also compute the corresponding probabilities $p(F_{\rm align},k$) $\propto$ exp[ln\,${\mathcal{L}}(F_{\rm align},k$)] to determine the overall uncertainties in and correlation between the model parameters.

\begin{figure}[t!]
\includegraphics[scale = 0.6]{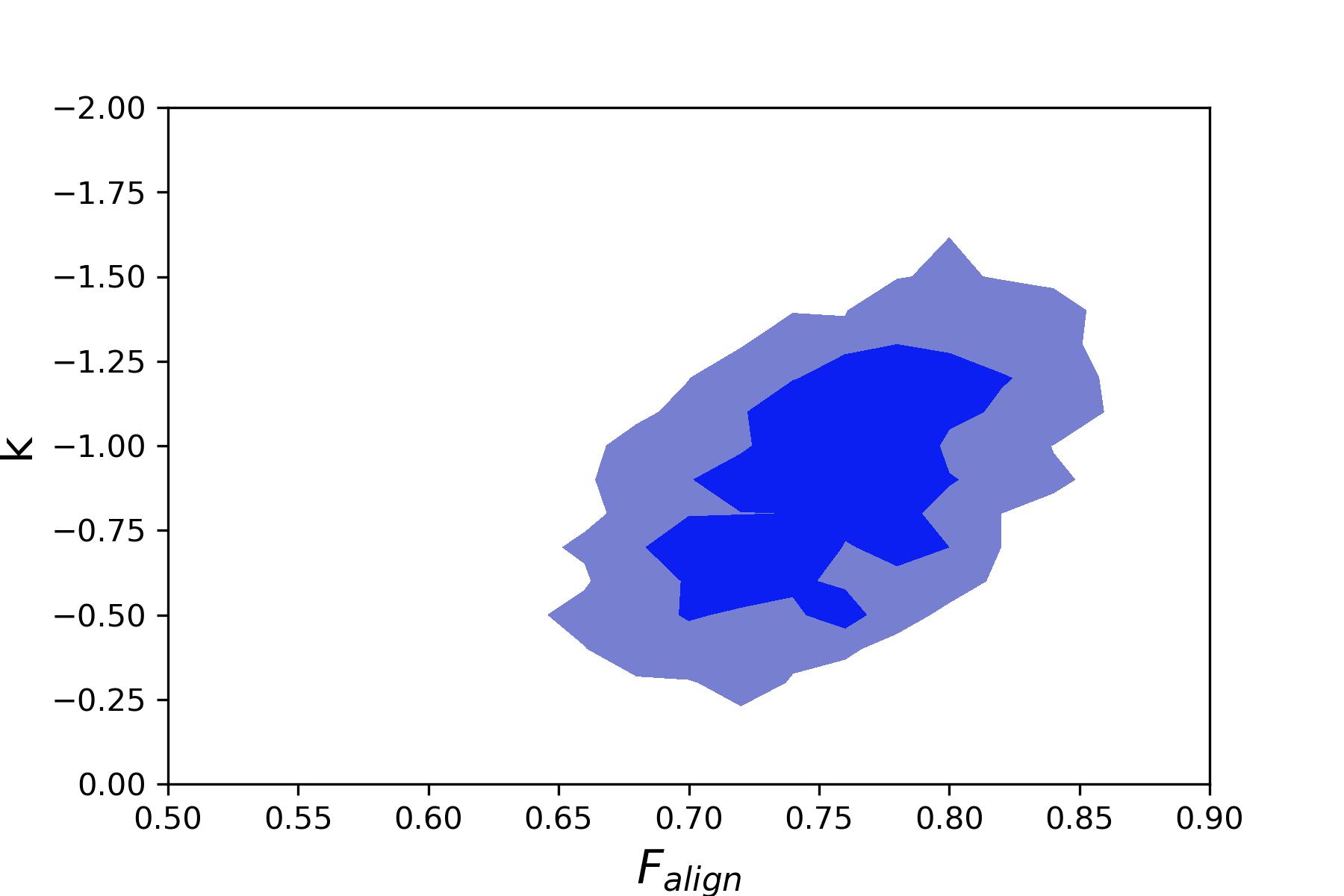}
\caption{The 1$\sigma$ (dark blue) and 2$\sigma$ (light blue) 2D confidence intervals for our two model parameters $F_{\rm align}$ and $k$. The 1D confidence intervals are $F_{\rm align}$ = 0.75\,$\pm$\,0.05 and $k$ = $-$0.9\,$\pm$\,0.4, corresponding to an average spin reduction factor of $\langle S_{\rm align} \rangle$ = 0.43\,$\pm$\,0.04.  } 
\label{fig:hmapfull}
\end{figure}

We display the distributions of $r_{\rm vsini}$ according to our best-fit model alongside the data in Fig.~\ref{fig:CDFbfit}. We also present the 2D confidence intervals for $F_{\rm align}$ and $k$ in Fig.~\ref{fig:hmapfull}. The fraction of astrometric binaries that exhibit spin-orbit alignment is $F_{\rm align}$ = 0.75\,$\pm$\,0.05, consistent with our initial estimate of $F_{\rm align}$ $\approx$ 70\% in section~\ref{sec:ecc}. Similarly, we measure $k$ = $-$0.9\,$\pm$\,0.4, which falls within our expected range of $-$3.5 $<$ $k$ $<$ 0. Our measured coefficient $k$ = $-$0.9\,$\pm$\,0.4 corresponds to an average spin reduction factor of  $\langle S_{\rm align} \rangle$ = 0.43\,$\pm$\,0.04. The slow-rotator blue sequence exhibits on average 35\%\,-\,55\% the rotation rate of the fast-rotator red sequence \citep[][see section \ref{sec:SlowRotators}]{Marino2018,Bastian2020,Wang2022}, which is remarkably consistent with our result of $\langle S_{\rm align} \rangle$ = 0.43\,$\pm$\,0.04.

\section{Conclusions}
\label{sec:Conclusions}

We summarize our main results as follows:

\begin{enumerate}
\item We created a catalog of 917 {\it Gaia} astrometric binaries with distances $d$ $<$ 500 pc, orbital periods $P$ = 100\,-\,3,000 days ($a$ = 0.5\,-\,5 au), primary spectral types B8-F1\,IV/V, and measured projected rotational velocities $v$\,sin\,$i$ from the {\it Gaia} RVS instrument and/or ground-based spectra. 
\item For the 207 systems with both {\it Gaia} and ground-based $v$\,sin\,$i$, we find good overall agreement and only minor $\approx$\,10\% systematic offsets, depending on the spectroscopic survey.
\item The early-type primaries in astrometric binaries with face-on orbits display substantially smaller $v$\,sin\,$i$ than those in edge-on orbits, signifying substantial spin-orbit alignment at the 6$\sigma$ confidence level. 
\item The early-type primaries in astrometric binaries are spinning more slowly than all stars of the same spectral type at the 8$\sigma$ confidence level and therefore comprise the slow-rotator population in the observed bimodal rotational velocity distribution.
\item Utilizing a Monte Carlo population synthesis technique, we determine that 75\%\,$\pm$\,5\% of early-type astrometric binaries exhibit spin-orbit alignment and that their primaries are rotating on average at 43\%\,$\pm$\,4\% the rate of all stars of the same spectral type. The measured average reduction in primary spin matches the observed ratio between the mean rotational velocities of the slow-rotator versus fast-rotator populations.
\item The 75\% of early-type astrometric binaries that exhibit spin-orbit alignment and reduced primary spin probably formed through disk fragmentation, inward disk migration, and modest circumbinary disk accretion. Compared to single stars or components in wide binaries, the pre-MS primaries in binaries across $a$ = 0.5\,-\,10~au likely accreted a larger fraction of their final mass at earlier times while still below the Kraft break. In this regime, magnetic braking could efficiently transfer angular momentum from the primary's spin though the disk and into the orbit. Even after the pre-MS primaries contracted and evolved above the Kraft break, the secondary accreted most of the remaining circumbinary disk mass, quenching angular momentum flow to the primary. 
\item The degree of spin-orbit alignment increases with primary mass across $M_1$ = 1.5\,-\,3.0\,\Msun, suggesting that the propensity for disk fragmentation and inward disk migration increases with final primary mass, which is consistent with model predictions.
\item The primaries in astrometric binaries with { small mass ratios $q$~$<$~0.32} display even slower spins and a higher degree of spin-orbit alignment. Disk fragmentation and inward disk migration can more readily produce close binaries with extreme mass ratios. Moreover, low-mass companions accrete a larger fraction of the final circumbinary disk mass and angular momentum, resulting in primaries with even slower spins.
\item The 30\% of astrometric binaries with large eccentricities $e$ $>$ 0.4 display neither spin-orbit alignment nor reduced spins. The majority of these eccentric binaries likely formed via core fragmentation on large scales followed by orbital decay via dynamical friction.
\item The nearly circular astrometric binaries with $e$~$<$~0.2 display even stronger spin-orbit alignment and slower spins. Although significant circumbinary disk accretion can dampen the eccentricities and align the orbits to the spins, such a process also drives the binary mass ratio toward unity, contrary to \#8 above. Current hydrodynamic simulations cannot yet produce close early-type binaries with small eccentricities, small mass ratios, and aligned spin-orbits that derive from core fragmentation. It is more plausible that close early-type binaries with small mass ratios are born with their small eccentricities and reduced, aligned spins via disk fragmentation, inward disk migration, and only modest circumbinary disk accretion. Large-scale magnetohydrodynamic simulations with higher resolution are necessary to test our interpretation of the various statistically significant trends.
\item Early-type primaries in binaries with $a$ = 0.5\,-\,10~au, $q$~$<$~0.8, and $e$~$<$~0.4 comprise the slow-rotator population. The binary fraction within this parameter space scales as 6\%($M_1$/\Msun)$^{0.6}$, which quantitatively matches the observed increase in the slow-rotator fraction with stellar mass.

 
\end{enumerate}

We thank Chen Wang for illuminating discussions on the slow-rotator population.


\appendix

\begin{adjustbox}{angle=90}
\begin{tabular}{c}
Table A1: Final Astrometric Binary Sample (first hour RA = 0$^{\circ}$\,-\,15$^{\circ}$ in printed version)
\end{tabular}
\end{adjustbox}
\label{tab:RAlist}
\def\arraystretch{2.1}
\begin{adjustbox}{angle=90}
\setlength{\tabcolsep}{4pt}

\resizebox{25cm}{!}{
\begin{tabular}{|rrrcccccccccccccccccccccccccccc|} 
\hline
RA\,(\deg)~ & Dec\,(\deg)~ & {\it Gaia} DR3 ID~~~~~ &Name&$\varpi$\,(mas) &$d$\,(pc) &G &$A_G$ & $M_{\rm G}$ &(BP-RP) & (BP-RP)$_{\rm o}$ &$\Delta M_{\rm G}$&P\,(d) & e & i\,(\deg) & f$_{\rm M}$\,(\Msun)& $M_1$ & $q$ & GSP~$T_{\rm eff}$\,(K) & GSP~log\,$g$ &GSP~ST &$v_{\rm broad}$ &$\langle v$\,sin\,$i\rangle_{\rm ST}$ & $r_{\rm vsini}$ & $T_{\rm eff}$\,(K) &log\,$g$&ST &$v\,{\rm sin}\,i$ &$\langle v\,{\rm sin}\,i\rangle_{\rm ST}$ & $r_{\rm vsini}$ & Ref \\\midrule
0.070135&59.050897&423153204646998912&TYC 3664-247-1&2.08&482&10.77&0.65&1.70&0.51&0.18&0.58&702&0.10&31&0.0188&1.73&>0.50&8160&4.29&A5V&52&97&0.53&&&A5V&55&120&0.46&6\\
0.372107 &-62.045775 &4904779744963317632 &CD-62 1473 &2.63&380&9.93&0.04&1.99&0.42&0.40&0.22&201&0.36&118&0.0162&1.71&<0.32&7225&3.93&F0V&&&&7041&3.95&F1V&50&74&0.68&3\\
1.323714 &54.463240 &420304709312762880 &TYC 3656-1660-1 &2.38&420&10.95&0.29&2.54&0.54&0.40&-0.06&669&0.20&128&0.0595&1.58&0.32-0.50&7156&4.09&F0V&41&90&0.45&&&&&&&\\
1.950457 &21.745437 &2847003051649951360 &BD+20 4 &3.65&274&9.81&0.08&2.54&0.33&0.29&-0.78&773&0.24&49&0.0219&1.97&<0.32&7599&3.87&A8V&34&98&0.34&&&&&&&\\
1.981364 &49.791271 &393740787563466880 &TYC 3254-1632-1 &2.71&369&10.55&0.32&2.40&0.38&0.22&-0.04&839&0.06&57&0.0351&1.68&0.32-0.50&7911&4.27&A6V&63&99&0.63&&&&&&&\\
2.187406 &-52.328286 &4972421077532967168 &HD 429 &4.70&213&8.58&0.08&1.86&0.44&0.40&0.30&557&0.04&69&0.0087&1.73&>0.32&7077&3.84&F1V&52&76&0.68&7202&3.99&F0V&66&91&0.72&3\\
2.385776 &47.974898 &393159008472127872 &BD+47 12B &2.61&383&10.77&0.21&2.65&0.56&0.46&-0.07&190&0.29&87&0.0433&1.53&0.32-0.50&6939&4.05&F1V&18&76&0.22&&&&&&&\\
2.856673 &-41.276682 &4996208294340751104 &CD-41 30 &2.13&471&10.44&0.06&2.02&0.49&0.46&0.28&423&0.11&107&0.0178&1.66&<0.32&6924&3.90&F1V&59&76&0.77&&&&&&&\\
3.328966 &57.206220 &422383237268603264 &TYC 3660-1821-1 &2.24&446&10.11&0.59&1.27&0.32&0.03&0.24&503&0.61&133&0.0462&2.01&0.32-0.50&9280&4.03&A1V&&&&8996&4.20&A2V&205&122&1.68&4\\
3.679295 &-54.756328 &4922973771889699456 &CD-55 36&2.04&490&10.47&0.05&1.96&0.31&0.28&0.29&780&0.09&128&0.0199&1.74&<0.32&7805&4.10&A7V&99&99&1.00&7468&4.16&A9V&125&98&1.28&3 \\
3.906941 &-14.636101 &2417064215296987136 &BD-15 35 &22.05&488&10.43&0.00&1.99&0.38&0.39&0.01&1036&0.20&17&0.0212&1.81&<0.32&7339&3.78&A9IV&25&98&0.25&7018&3.92&F1V&36&74&0.49&3\\
5.337729 &46.711986 &392061249190173312 &TYC 3247-936-1 &2.09&479&10.34&0.23&1.71&0.53&0.42&0.21&1001&0.20&165&0.0924&1.83&0.32-0.50&7078&3.77&F1IV&60&99&0.60&&&&&&&\\
5.566180 &55.368862 &421372373765588480 &TYC 3657-1737-1 &3.71&269&10.05&0.14&2.76&0.39&0.32&-0.15&555&0.18&75&0.0378&1.59&0.32-0.50&7435&4.26&A9V&90&98&0.91&&&&&&&\\
5.955202 &-5.637617 &2431838627916768256 &HD 1959&3.75&267&10.00&&&0.44&&&496&0.03&28&0.0167&&&&&A8V&29&98&0.29&&&&&&&\\
6.659723 &44.040468 &382697945607189248 &HD 2255 &6.68&150&8.97&0.13&2.96&0.50&0.43&-0.38&138&0.12&78&0.0306&1.52&0.32-0.50&7018&4.08&F1V&66&76&0.86&&&&&&&\\
6.783298 &-8.936009 &2426822174834338304 &HD 2334 &2.10&477&10.23&0.05&1.79&0.41&0.39&0.29&808&0.68&34&0.0441&1.77&0.32-0.50&7458&3.98&A9V&&&&6909&3.81&F1V&75&74&1.02&3 \\
6.811769 &-59.212000 &4906274977697877632 &HD 2385 &4.56&219&8.98&0.01&2.26&0.45&0.44&0.19&1159&0.81&92&0.0106&1.58&<0.32&6950&3.98&F1V&26&76&0.33&7021&4.03&F1V&45&74&0.61&3 \\
7.687790 &54.404326 &418085349398538752 &BD+53 80 &2.46&406&9.69&0.32&1.33&0.41&0.25&0.45&242&0.26&86&0.0439&1.95&>0.50&7803&3.93&A7V&79&99&0.80&&&&&&&\\
8.785423 &-57.387320 &4908164247912248064 &TYC 8468-220-1 &22.51&399&10.80&0.18&2.61&0.44&0.35&-0.06&677&0.47&122&0.0356&1.59&0.32-0.50&7318&4.19&A9V&10&98&0.08&&&&&&&\\
9.90 &-59.513576 &4906782230515474176 &HD 3751 &3.46&289&9.47&0.08&2.09&0.33&0.29&-0.43&709&0.68&157&0.0488&2.02&0.32-0.50&7589&3.81&A8V&228&98&2.32&&&&&&&\\
10.351935 &42.885461 &387433164229245312 &BD+42 145 &3.25&308&10.27&0.22&2.60&0.54&0.43&0.13&487&0.20&81&0.0377&1.45&0.32-0.50&7039&4.17&F1V&28&76&0.36&&&&&&&\\
10.707604 &56.506860 &424541819113465344 &TYC 3663-2392-1 &22.28&438&10.49&0.53&1.76&0.40&0.14&-0.10&731&0.03&144&0.0145&1.96&<0.32&8430&4.03&A3V&76&104&0.73&&&&&&&\\
11.462197 &-65.165478 &4708826088331213568 &HD 4473 &2.59&386&9.53&0.25&1.35&0.16&0.03&0.17&274&0.27&52&0.0459&2.01&0.32-0.50&9776&4.16&A0V&288&125&2.30&8348&4.05&A4V&321&119&2.71&3\\
12.290008 &20.560933 &2801417020388676992 &TYC 1194-391-1 &33.25&307&9.95&0.21&2.30&0.43&0.32&-0.32&719&0.50&20&0.0586&1.88&0.32-0.50&7565&4.04&A8V&&&&7268&3.83&F0V&114&82&1.39&2\\
12.323751 &-40.514603 &4999172440249827456 &CD-41 211 &3.09&323&10.06&0.06&2.45&0.46&0.43&0.05&704&0.10&66&0.0320&1.56&0.32-0.50&7041&4.05&F1V&89&76&1.16&&&&&&&\\
13.017756 &55.993499 &423682997453694080 &TYC 3659-875-1 &2.18&458&10.62&0.43&1.89&0.43&0.22&0.14&934&0.29&134&0.0687&1.83&0.32-0.50&7912&3.86&A6V&194&99&1.96&7987&4.36&A6V&202&92&2.19&2\\
13.682239 &-35.522139 &5002580582699312000 &CD-36 315 &2.96&337&10.33&0.11&2.59&0.51&0.46&0.13&512&0.41&134&0.0098&1.47&<0.32&6936&4.12&F1V&73&76&0.95&&&&&&&\\
14.300333 &59.524156 &426045023307316224 &TYC 3680-605-1 &2.64&378&10.20&0.60&1.71&0.54&0.24&0.03&509&0.71&48&0.0532&1.96&0.32-0.50&&&A6V&205&99&2.07&7840&3.93&A6V&197&92&2.14&2 \\
14.913646 &-11.933699 &2469440894794921088 &HD 5823 &3.01&332&9.92&0.09&2.22&0.39&0.34&0.04&703&0.15&87&0.0360&1.74&0.32-0.50&7342&4.05&A9V&26&98&0.26&&&&&&&\\
\hline
\end{tabular}}
\end{adjustbox}
\begin{adjustbox}{angle=90}
\begin{tabular}{c}
\footnotesize{References: 1\,-\,\citet{Glebocki2005}; 2\,-\,\citet{Xiang2022}; 3\,-\,\citet{Steinmetz2020}; 4\,-\,\citet{Sun2021}; 5\,-\,\citet{Buder2021}; 6\,-\,\citet{Jonsson2020}}
\end{tabular}
\end{adjustbox}

\newpage

\bibliographystyle{apj}                       
\bibliography{moe_biblio}

\end{document}